\documentclass[preprint,prd,aps,showpacs,showkeys,nofootinbib]{revtex4}
\usepackage{graphicx}
\begin{document}
\title{The extended BLMSSM with a 125 GeV Higgs boson and dark matter}

\author{Shu-Min Zhao$^1$\footnote{email:zhaosm@hbu.edu.cn}, Tai-Fu Feng$^{1}$\footnote{email:fengtf@hbu.edu.cn},
Guo-Zhu Ning$^{1}$\footnote{ninggz@hbu.edu.cn}, Jian-Bin Chen$^{2}$\footnote{chenjianbin@tyut.edu.cn},
Hai-Bin Zhang$^{1}$\footnote{hbzhang@hbu.edu.cn}, Xing Xing Dong$^{1}$}

\affiliation{$^1$ Department of Physics, Hebei University, Baoding 071002, China,\\
$^2$ College of Physics and Optoelectronic Engineering,
Taiyuan University of Technology, Taiyuan 030024, China}

\date{\today}

\begin{abstract}
To extend the BLMSSM, we not only add exotic Higgs superfields $(\Phi_{NL},\varphi_{NL})$ to make the exotic lepton heavy, but also introduce the
superfields($Y$,$Y^\prime$) having couplings with lepton and exotic lepton at tree level. The obtained model is called as EBLMSSM, which has
difference from BLMSSM especially for the exotic slepton(lepton) and exotic sneutrino(neutrino). We deduce the mass matrices and the needed couplings in this model.
To confine the parameter space, the Higgs boson mass $m_{h^0}$ and the processes $h^0\rightarrow \gamma\gamma$, $h^0\rightarrow VV, V=(Z,W)$ are
studied in the EBLMSSM. With the assumed parameter space, we obtain reasonable numerical results according to
data on Higgs from ATLAS and CMS. As a cold dark mater candidate, the relic density for the lightest mass eigenstate of $Y$ and $Y'$ mixing is also studied.
\end{abstract}


\maketitle
\section{introduction}

The total lepton number (L) and baryon number (B) are good symmetries because neutrinoless double beta decay or proton decay has not
been observed. In the standard model(SM), L and B are global symmetries\cite{BLWending}. However, the individual lepton numbers $L_i=L_e,~L_\mu,~L_\tau$ are not
exact symmetries at the electroweak scale because of the neutrino oscillation and the neutrinos with tiny masses\cite{neutrinomass}. In the Universe, there is
matter-antimatter asymmetry, then the baryon number must be broken.

With the detection of the light Higgs $h^0(m_h^0=125.1{\rm GeV})$\cite{higgs125}, the SM succeeds greatly and the Higgs mechanism is compellent.
Beyond the SM, supersymmetry\cite{SUSY} provides a possibility to understand the light Higgs. The minimal supersymmetric
extension of the SM(MSSM)\cite{MSSM} is one of the favorite models, where the light Higgs mass at tree level is $m_{h}^{tree}=m_Z|\cos2\beta|$ \cite{BLfirst}. The one loop corrections to
Higgs mass mainly come from fermions and sfermions, that depend on the virtual particle masses and the couplings with the Higgs.

There are many papers about the gauged B and L models, although most of them are non-supersymmetric \cite{BLnoSYSY}.
Extending MSSM with the local gauged B and L, one obtains the so called BLMSSM,
which was proposed by the authors in Ref.\cite{BLfirst}. The proton remains stable, as B and L are broken at the TeV scale.
Therefore, a large desert between the electroweak scale and grand unified scale is not necessary.
In BLMSSM, the baryon number is changed by one unit, at the same time the lepton number is broken in an even number.
 R-parity in BLMSSM is not conserved, and it can explain the matter-antimatter asymmetry in the Universe.  There are some works for Higgs and
 dark matters\cite{dark1}
in the BLMSSM\cite{darkM,TFBL}. In the framework of BLMSSM, the light Higgs mass and the decays $h^0\rightarrow \gamma\gamma$ and $h^0\rightarrow VV, V=(Z,W)$ are studied in
 our previous work\cite{TFBL}. Some lepton flavor violating processes and CP-violating processes are researched with the new parameters in BLMSSM\cite{zhaolepton}.

In BLMSSM, the exotic leptons are not heavy, because their masses just have relation with the parameters $Y_{e_4}\upsilon_d,~Y_{e_5}\upsilon_u$.
 Here $\upsilon_u$ and $\upsilon_d$ are the vacuum expectation values(VEVs) of two Higgs doublets $H_u$ and $H_d$.
 In general, the Yukawa couplings $Y_{e_4}$ and $Y_{e_5}$ are not large
 parameters, so the exotic lepton masses are around 100 GeV. The light exotic leptons may lead to
  that the BLMSSM is excluded by high energy physics experiments
 in the future. To obtain heavy exotic leptons, we add two exotic Higgs superfields to the BLMSSM, and they are
 SU(2) singlets $\Phi_{NL}$ and $\varphi_{NL}$, whose VEVs are $\upsilon_{NL}$ and
 $\bar{\upsilon}_{NL}$\cite{cd750}. The exotic leptons and the superfields
 $\Phi_{NL},\varphi_{NL}$ have Yukawa couplings, then $\upsilon_{NL}$ and $\bar{\upsilon}_{NL}$ give
 contributions to the diagonal elements of the exotic lepton mass matrix.
 So the exotic leptons turn heavy and  should be unstable. In the end, the super fields $Y$ and $Y'$ are also introduced.
 At tree level, there are couplings for lepton-exotic lepton-$Y(Y')$.
 It is appealing that this extension of BLMSSM produces some new cold dark matter candidates,
 such as the lightest mass egeinstate of $Y$ and $Y'$ mixing. The four-component
 spinor $\tilde{Y}$ is made up of the superpartners of $Y$ and $Y'$.
 In this extended BLMSSM(EBLMSSM), we study the lightest CP even Higgs mass with the one loop corrections.
 The Higgs decays $h^0\rightarrow \gamma\gamma$ and $h^0\rightarrow VV, ~V=(Z, W)$ are also calculated here.
 Supposing the lightest mass eigenstate of $Y$ and $Y'$ mixing as a cold dark matter candidate, we study the relic density.

After this introduction, in Section 2, we introduce the EBLMSSM in detail,
including the mass matrices and the couplings different from those in the BLMSSM.
The mass of the lightest CP-even Higgs $h^0$ is deduced in the Section 3. The Section 4 is used to give the
formulation of the Higgs decays  $h^0\rightarrow \gamma\gamma$, $h^0\rightarrow VV, ~V=(Z, W)$ and dark matter relic density.
The corresponding numerical results are computed in Section 5.
The last section is used for the
 discussion and conclusion.

\section{Extend the BLMSSM}
The local gauge group of the BLMSSM \cite{BLfirst} is
$SU(3)_{C}\otimes SU(2)_{L}\otimes U(1)_{Y}\otimes U(1)_{B}\otimes U(1)_{L}$. In the BLMSSM, the exotic lepton
masses are obtained from the Yukawa couplings with the two Higgs doublets $H_u$ and $H_d$. The VEVs
 of $H_u$ and $H_d$ are $\upsilon_u$ and $\upsilon_d$ with the relation $\sqrt{\upsilon_u^2+\upsilon_d^2}=\upsilon\sim 250$ GeV.
Therefore, the exotic lepton masses are not very heavy, though they can satisfy the experiment bounds at present.
In the future, with the development of high energy experiments, the experiment bounds for the exotic lepton masses
 can improve in a great possibility. Therefore, we introduce the exotic Higgs superfields $\Phi_{NL}$ and $\varphi_{NL}$
 with nonzero VEVs to make the exotic lepton heavy. The heavy exotic leptons should be unstable,
 then the superfields $Y,Y'$ are introduced accordingly. These introduced superfields lead to tree level couplings for lepton-exotic lepton-$Y(Y')$.

In EBLMSSM, we show the superfields in the Table I.

\begin{table}
\caption{ The super fields in the extended BLMSSM (EBLMSSM)}
\begin{tabular}{|c|c|c|c|c|c|}
\hline
Superfields & $SU(3)_C$ & $SU(2)_L$ & $U(1)_Y$ & $U(1)_B$ & $U(1)_L$\\
\hline
$\hat{Q}_i$ & 3 & 2 & 1/6 & $1/3$ & 0 \\
\hline
$\hat{u}^c_i$ & $\bar{3}$ & 1 & -2/3 & -$1/3$ & 0\\
\hline
$\hat{d}^c_i$ & $\bar{3}$ & 1 & 1/3 & -$1/3$ & 0 \\
\hline
$\hat{L}_i$ & 1 & 2 & -1/2 & 0 & $1$ \\
\hline
$\hat{e}^c_i$ & 1 & 1 & 1 & 0 & -$1$ \\
\hline
$\hat{N}^c_i$ & 1 & 1 & 0 & 0 & -$1$\\
\hline
$\hat{Q}_4$ & 3 & 2 & 1/6 & $B_4$ & 0 \\
\hline
$\hat{U}^c_4$ & $\bar{3}$ & 1 & -2/3 & -$B_4$ & 0 \\
\hline
$\hat{D}^c_4$ & $\bar{3}$ & 1 & 1/3 & -$B_4$ & 0 \\
\hline
$\hat{Q}_5^c$ & $\bar{3}$ & 2 & -1/6 & -$(1+B_4)$ & 0 \\
\hline
$\hat{U}_5$ & $3$ & 1 & 2/3 &  $1 + B_4$ & 0 \\
\hline
$\hat{D}_5$ & $3$ & 1 & -1/3 & $1 + B_4$ & 0 \\
\hline
$\hat{L}_4$ & 1 & 2 & -1/2 & 0 & $L_4$ \\
\hline
$\hat{E}^c_4$ & 1 & 1 & 1 & 0 & -$L_4$ \\
\hline
$\hat{N}^c_4$ & 1 & 1 & 0 & 0 & -$L_4$ \\
\hline
$\hat{L}_5^c$ & 1 & 2 & 1/2 & 0 & -$(3 + L_4)$ \\
\hline
$\hat{E}_5$ & 1 & 1 & -1 & 0 & $3 + L_4$ \\
\hline
$\hat{N}_5$ & 1 & 1 & 0 & 0 & $3 + L_4$ \\
\hline
$\hat{H}_u$ & 1 & 2 & 1/2 & 0 & $0$ \\
\hline
$\hat{H}_d$ & 1 & 2 & -1/2 & 0 & $0$ \\
\hline
$\hat{\Phi}_B$ & 1 & 1 & 0 & 1 & 0 \\
\hline
$\hat{\varphi}_B$ & 1 & 1 & 0 & -1 & 0 \\
\hline
$\hat{\Phi}_L$ & 1 & 1 & 0 & 0 & -2 \\
\hline
$\hat{\varphi}_L$ & 1 & 1 & 0 & 0 & 2 \\
\hline
$\hat{\Phi}_{NL}$ & 1 & 1 & 0 & 0 & -3 \\
\hline
$\hat{\varphi}_{NL}$ & 1 & 1 & 0 & 0 & 3\\
\hline
$\hat{X}$ & 1 & 1 & 0 & $2/3 + B_4$ & 0 \\
\hline
$\hat{X'}$ & 1 & 1 & 0 & $-(2/3 + B_4)$ & 0 \\
\hline
$Y$ & 1 & 1 & 0 & 0 & $2+L_4$
\\ \hline
$Y'$ & 1 & 1 & 0 & 0 & $-(2+L_4)$ \\
\hline
\end{tabular}
\label{quarks}
\end{table}

The superpotential of EBLMSSM is shown here
\begin{eqnarray}
&&{\cal W}_{{EBLMSSM}}={\cal W}_{{MSSM}}+{\cal W}_{B}+{\cal W}_{L}+{\cal W}_{X}+{\cal W}_{Y}\;,
\nonumber\\
&&{\cal W}_{L}=\lambda_{L}\hat{L}_{4}\hat{L}_{5}^c\hat{\varphi}_{NL}+\lambda_{E}\hat{E}_{4}^c\hat{E}_{5}
\hat{\Phi}_{NL}+\lambda_{NL}\hat{N}_{4}^c\hat{N}_{5}\hat{\Phi}_{NL}
+\mu_{NL}\hat{\Phi}_{NL}\hat{\varphi}_{NL}\nonumber\\&&\hspace{1.2cm}+Y_{{e_4}}\hat{L}_{4}\hat{H}_{d}\hat{E}_{4}^c+Y_{{\nu_4}}\hat{L}_{4}\hat{H}_{u}\hat{N}_{4}^c
+Y_{{e_5}}\hat{L}_{5}^c\hat{H}_{u}\hat{E}_{5}+Y_{{\nu_5}}\hat{L}_{5}^c\hat{H}_{d}\hat{N}_{5}
\nonumber\\
&&\hspace{1.2cm}
+Y_{\nu}\hat{L}\hat{H}_{u}\hat{N}^c+\lambda_{{N^c}}\hat{N}^c\hat{N}^c\hat{\varphi}_{L}
+\mu_{L}\hat{\Phi}_{L}\hat{\varphi}_{L}\;,
\nonumber\\&&
{\cal W}_{Y}=\lambda_4\hat{L}\hat{L}_{5}^c\hat{Y}+\lambda_5\hat{N}^c\hat{N}_{5}\hat{Y}^\prime
+\lambda_6\hat{E}^c\hat{E}_{5}\hat{Y}^\prime+\mu_{Y}\hat{Y}\hat{Y}^\prime\;.
\end{eqnarray}
${\cal W}_{{MSSM}}$ is the superpotential of MSSM. ${\cal W}_{B}$ and ${\cal W}_{X}$ are same as the terms in BLMSSM\cite{TFBL}.
$W_Y$ includes the terms beyond BLMSSM, and they include the couplings of lepton-exotic lepton-$Y$($l^I-L'-Y$). Therefore, the heavy
exotic leptons can decay to leptons and mass eigenstates of $Y$ and $Y^\prime$ mixing whose lighter one can be  a dark matter candidate.
From $W_Y$, one can also obtain the coupling of lepton-exotic slepton-$\tilde{Y}$ ($l^I-\tilde{L}'-\tilde{Y}$),
 where $\tilde{Y}$ is the four component spinor composed by the superpartners of $Y$ and $Y'$. 
 The new couplings of $l^I-L'-Y$ and $l^I-\tilde{L}'-\tilde{Y}$ can
give one loop corrections to lepton anormal magnetic dipole moment(MDM). They may compensate the deviation between the experiment
value and SM prediction for muon MDM. The parameter $\mu_Y$ can be
complex number with non-zero imaginary part, which is a new source of CP-violating.
Therefore, the both new couplings produce one loop diagrams contributing to the lepton
electric dipole moment(EDM). Further more, if $\lambda_4$ in $\lambda_4\hat{L}\hat{L}_{5}^c\hat{Y}$ is a matrix and has
non-zero elements relating with lepton flavor, this term can enhance the lepton flavor violating effects.
In the whole, ${\cal W}_{Y}$ enriches the lepton physics to a certain degree, and these subjects will be researched in our latter works.

Because of the introduction of the superfields $\Phi_{NL},\varphi_{NL}, Y$ and $Y'$, the soft breaking terms are
written as
\begin{eqnarray}
&&{\cal L}_{{soft}}^{EBLMSSM}={\cal L}_{{soft}}^{BLMSSM}
-m_{{\Phi_{NL}}}^2\Phi_{NL}^*\Phi_{NL}
-m_{{\varphi_{NL}}}^2\varphi_{NL}^*\varphi_{NL}
+(A_{{LL}}\lambda_{L}\tilde{L}_{4}\tilde{L}_{5}^c\varphi_{NL}\nonumber\\&&\hspace{2.0cm}
+A_{{LE}}\lambda_{E}\tilde{e}_{4}^c\tilde{e}_{5}\Phi_{NL}
+A_{{LN}}\lambda_{NL}\tilde{\nu}_{4}^c\tilde{\nu}_{5}\Phi_{NL}
+B_{NL}\mu_{NL}\Phi_{NL}\varphi_{NL}+h.c.)\nonumber\\&&\hspace{2.0cm}+(
A_4\lambda_4\tilde{L}\tilde{L}_{5}^cY+A_5\lambda_5\tilde{N}^c\tilde{\nu}_{5}Y^\prime
+A_6\lambda_6\tilde{e}^c\tilde{e}_{5}Y^\prime+B_{Y}\mu_{Y}YY^\prime+h.c.).
\label{soft-breaking}
\end{eqnarray}
Here ${\cal L}_{{soft}}^{BLMSSM}$ is the soft breaking terms of BLMSSM, whose concrete form is in our previous work\cite{TFBL}.
The $SU(2)_L$ doublets $H_{u},H_{d}$ acquire the nonzero VEVs $\upsilon_{u},\upsilon_{d}$.
The $SU(2)_L$ singlets $\Phi_{B},\varphi_{B},\Phi_{L},\varphi_{L},\Phi_{NL},\varphi_{NL}$
obtain the nonzero VEVs $\upsilon_{{B}},\overline{\upsilon}_{{B}},\upsilon_{L},\;\overline{\upsilon}_{L},
\upsilon_{NL},\;\overline{\upsilon}_{NL}$ respectively.

\begin{eqnarray}
&&H_{u}=\left(\begin{array}{c}H_{u}^+\\{1\over\sqrt{2}}\Big(\upsilon_{u}+H_{u}^0+iP_{u}^0\Big)\end{array}\right),
~~~~~~
H_{d}=\left(\begin{array}{c}{1\over\sqrt{2}}\Big(\upsilon_{d}+H_{d}^0+iP_{d}^0\Big)\\H_{d}^-\end{array}\right),
\nonumber\\
&&\Phi_{B}={1\over\sqrt{2}}\Big(\upsilon_{B}+\Phi_{B}^0+iP_{B}^0\Big),~~~~~~~~~~~
\varphi_{B}={1\over\sqrt{2}}\Big(\overline{\upsilon}_{B}+\varphi_{B}^0+i\overline{P}_{B}^0\Big),
\nonumber\\
&&\Phi_{L}={1\over\sqrt{2}}\Big(\upsilon_{L}+\Phi_{L}^0+iP_{L}^0\Big),~~~~~~~~~~~~
\varphi_{L}={1\over\sqrt{2}}\Big(\overline{\upsilon}_{L}+\varphi_{L}^0+i\overline{P}_{L}^0\Big),
\nonumber\\
&&\Phi_{NL}={1\over\sqrt{2}}\Big(\upsilon_{NL}+\Phi_{NL}^0+iP_{NL}^0\Big),~~~~
\varphi_{NL}={1\over\sqrt{2}}\Big(\overline{\upsilon}_{NL}+\varphi_{NL}^0+i\overline{P}_{NL}^0\Big).
\end{eqnarray}
Here, we define $\tan\beta=\upsilon_u/\upsilon_d,~\tan\beta_B=\bar{\upsilon}_B/\upsilon_B,~\tan\beta_L=\bar{\upsilon}_L/\upsilon_L$ and
$\tan\beta_{NL}=\bar{\upsilon}_{NL}/\upsilon_{NL}$.
The VEVs of the Higgs satisfy the following equations
\begin{eqnarray}
&&|\mu|^2-\frac{g_1^2+g_2^2}{8}(\upsilon_u^2-\upsilon_d^2)+m_{H_d}^2+Re[B\mu]\tan\beta=0,\\&&
|\mu|^2+\frac{g_1^2+g_2^2}{8}(\upsilon_u^2-\upsilon_d^2)+m_{H_u}^2+Re[B\mu]\cot\beta=0,\\&&
|\mu_B|^2+\frac{g_B^2}{2}(\upsilon_B^2-\bar{\upsilon}_B^2)+m_{\Phi_B}^2-Re[B_B\mu_B]\tan\beta_B=0,\\&&
|\mu_B|^2-\frac{g_B^2}{2}(\upsilon_B^2-\bar{\upsilon}_B^2)+m_{\varphi_B}^2-Re[B_B\mu_B]\cot\beta_B=0,\\&&
|\mu_L|^2-2g_L^2V_L^2+m_{\Phi_L}^2-Re[B_L\mu_L]\tan\beta_L=0,\label{VL1}\\&&
|\mu_L|^2+2g_L^2V_L^2+m_{\varphi_L}^2-Re[B_L\mu_L]\cot\beta_L=0,\label{VL2}\\&&
|\mu_{NL}|^2-3g_L^2V_L^2+m_{\Phi_{NL}}^2-Re[B_{NL}\mu_{NL}]\tan\beta_{NL}=0,\label{VL3}\\&&
|\mu_{NL}|^2+3g_L^2V_L^2
+m_{\varphi_{NL}}^2-Re[B_{NL}\mu_{NL}]\cot\beta_{NL}=0,\label{VL4}
\end{eqnarray}
with $V_L^2=\overline{\upsilon}^2_L-\upsilon^2_L+\frac{3}{2}(\overline{\upsilon}^2_{NL}-\upsilon^2_{NL})$.
Here, the Eqs.(\ref{VL1}) and (\ref{VL2}) are similar as the corresponding equations in BLMSSM, but Eqs.(\ref{VL1}) and (\ref{VL2})
have relation with the new parameters $\upsilon_{NL}$ and $\bar{\upsilon}_{NL}$. We obtain the new Eqs.(\ref{VL3}) and (\ref{VL4})
through $\frac{\partial V}{\partial \Phi_{NL}}$ and $\frac{\partial V}{\partial \varphi_{NL}}$, with $V$ denoting the Higgs scalar potential.

Here we deduce the mass matrices in the EBLMSSM.
Compared with BLMSSM, the superfields $\Phi_{NL}$ and $\varphi_{NL}$ are introduced and they give corrections to the
mass matrices of the slepton, sneutrino, exotic lepton, exotic neutrino, exotic slepton and exotic sneutrino.
That is to say, in EBLMSSM, the mass matrices of squark, exotic quark, exotic squark, baryon neutralino, MSSM neutralino,
$X$ and $\tilde{X}$ are same as those in the BLMSSM, and their concrete forms can be found in our previous works\cite{SM14JHEP,zhaoBL}.
Though the mass squared matrices of slepton and sneutrino in EBLMSSM  are different from those in BLMSSM,
we can obtain the slepton and sneutrino mass squared matrices in EBLMSSM easily just using the replacement
$\overline{\upsilon}^2_L-\upsilon^2_L\rightarrow V_L^2$ for the
BLMSSM results.

In the BLMSSM, the issue of Landau pole has been discussed in detail by the authors of Ref.\cite{BLfirst}.
Their conclusion is that there are no Landau poles at the low scale due to the new families.
In EBLMSSM, the parts of quark (squark), exotic quark (exotic squark) are same as those in BLMSSM.
Therefore, the Landau pole conditions for the Yukawa couplings of quark (squark), exotic quark (exotic squark) have same behaviors of BLMSSM.
The added superfields $(\Phi_{NL},\varphi_{NL}, Y, Y')$ do not have couplings with
the gauge fields of $SU(3)_C,SU(2)_L,U(1)_Y$ and $U(1)_B$. So the characters of gauge couplings $g_1,g_2,g_3$ and $g_B$ in BLMSSM and EBLMSSM are same.

The different parts between BLMSSM and EBLMSSM are the terms including $\Phi_{NL},\varphi_{NL}, Y$ and $ Y'$.
The new terms in the superpotential $\mathcal{W}_L$ are $\lambda_{L}\hat{L}_{4}\hat{L}_{5}^c\hat{\varphi}_{NL}
+\lambda_{E}\hat{E}_{4}^c\hat{E}_{5}
\hat{\Phi}_{NL}+\lambda_{NL}\hat{N}_{4}^c\hat{N}_{5}\hat{\Phi}_{NL}
+\mu_{NL}\hat{\Phi}_{NL}\hat{\varphi}_{NL}$ and they have corresponding relations with $\lambda_{Q}\hat{Q}_{4}\hat{Q}_{5}^c\hat{\Phi}_{B}+\lambda_{U}\hat{U}_{4}^c\hat{U}_{5}
\hat{\varphi}_{B}+\lambda_{D}\hat{D}_{4}^c\hat{D}_{5}\hat{\varphi}_{B}+\mu_{B}\hat{\Phi}_{B}\hat{\varphi}_{B}$ in $\mathcal{W}_B$ by the replacements
$\hat{L}_{4}\leftrightarrow \hat{Q}_{4}, \hat{L}^c_{5}\leftrightarrow \hat{Q}^c_{5}, \hat{E}^c_{4}\leftrightarrow \hat{U}^c_{4}, \hat{E}_{5}\leftrightarrow \hat{U}_{5}, \hat{N}^c_{4}\leftrightarrow \hat{D}^c_{4},\hat{N}_{5}\leftrightarrow \hat{D}_{5}, \hat{\Phi}_{NL}\leftrightarrow\hat{\varphi}_{B},\hat{\varphi}_{NL}\leftrightarrow\hat{\Phi}_{B}$.
The corresponding relations for  ${\cal W}_{Y}=\lambda_4\hat{L}\hat{L}_{5}^c\hat{Y}+\lambda_5\hat{N}^c\hat{N}_{5}\hat{Y}^\prime
+\lambda_6\hat{E}^c\hat{E}_{5}\hat{Y}^\prime+\mu_{Y}\hat{Y}\hat{Y}^\prime$ and ${\cal W}_{X}=\lambda_1\hat{Q}\hat{Q}_{5}^c\hat{X}+\lambda_2\hat{U}^c\hat{U}_{5}\hat{X}^\prime
+\lambda_3\hat{D}^c\hat{D}_{5}\hat{X}^\prime+\mu_{X}\hat{X}\hat{X}^\prime$ are obvious with $\hat{L}\leftrightarrow \hat{Q}, \hat{L}^c_{5}\leftrightarrow \hat{Q}^c_{5}, \hat{E}^c\leftrightarrow \hat{U}^c, \hat{E}_{5}\leftrightarrow \hat{U}_{5}, \hat{N}^c\leftrightarrow \hat{D}^c,\hat{N}_{5}\leftrightarrow \hat{D}_{5}, \hat{X}\leftrightarrow\hat{Y},\hat{X}^\prime\leftrightarrow\hat{Y}^\prime$.
From this analysis, the Landau pole conditions of gauge coupling $g_L$ and Yukawa couplings of exotic leptons
should possess similar peculiarities of gauge coupling $g_B$ and Yukawa couplings of exotic quarks. In conclusion, similar as BLMSSM, there are no Landau poles in EBLMSSM at the low scale because of the new families. The concrete study of Landau poles for the couplings should use renormalization group equation which is tedious, and we shall research this issue in our future work.

\subsection{The mass matrices of exotic lepton (slepton) and exotic neutrino (sneutrino) in EBLMSSM}
In BLMSSM, the exotic lepton masses are not heavy, because they obtain masses only from $H_u$ and $H_d$.
The VEVs of $\Phi_{NL}$ and $\varphi_{NL}$ are $\upsilon_{NL}$ and $\bar{\upsilon}_{NL}$, that can be large parameters.
So, the EBLMSSM exotic leptons are heavier than those in BLMSSM.

The mass matrix for the exotic leptons reads as
\begin{eqnarray}
&&-{\cal L}_{{e^\prime}}^{mass}=\left(\begin{array}{ll}\bar{e}_{{4R}}^\prime,&\bar{e}_{{5R}}^\prime\end{array}\right)
\left(\begin{array}{ll}-{1\over\sqrt{2}}\lambda_{L}\overline{\upsilon}_{NL},&{1\over\sqrt{2}}Y_{{e_5}}\upsilon_{u}\\
-{1\over\sqrt{2}}Y_{{e_4}}\upsilon_{d},&{1\over\sqrt{2}}\lambda_{E}\upsilon_{NL}
\end{array}\right)\left(\begin{array}{l}e_{{4L}}^\prime\\e_{{5L}}^\prime\end{array}\right)+h.c.
\label{ELmass}
\end{eqnarray}
Obviously, $\overline{\upsilon}_{NL}$ and $\upsilon_{NL}$ are the diagonal elements of the mass matrix in the Eq.(\ref{ELmass}).
It is easy to obtain heavy exotic lepton masses with large $\overline{\upsilon}_{NL}$ and $\upsilon_{NL}$. If we take $\overline{\upsilon}_{NL}$ and $\upsilon_{NL}$
as zero, the mass matrix is same as that in BLMSSM. In fact, our used values of $\overline{\upsilon}_{NL}$ and $\upsilon_{NL}$ are at TeV order, which produce
TeV scale exotic leptons. Heavy exotic leptons have strong adaptive capacity to the experiment bounds.
The exotic neutrinos are four-component spinors, whose  mass matrix is
\begin{eqnarray}
&&-{\cal L}_{{\nu^\prime}}^{mass}=\left(\begin{array}{ll}\bar{\nu}_{{4R}}^\prime,&\bar{\nu}_{{5R}}^\prime\end{array}\right)
\left(\begin{array}{ll}{1\over\sqrt{2}}\lambda_{L}\overline{\upsilon}_{NL},&-{1\over\sqrt{2}}Y_{{\nu_5}}\upsilon_{d}\\
{1\over\sqrt{2}}Y_{{\nu_4}}\upsilon_{u},&{1\over\sqrt{2}}\lambda_{NL}\upsilon_{NL}
\end{array}\right)\left(\begin{array}{l}\nu_{{4L}}^\prime\\\nu_{{5L}}^\prime\end{array}\right)+h.c.
\label{Nmass-matrix}
\end{eqnarray}
Similar as the exotic lepton condition, heavy exotic neutrinos are also gotten.

In BLMSSM, the exotic sleptons of 4 generation and 5 generation do not mix, and their mass matrices are both $2\times2$.
In EBLMSSM, the exotic sleptons of 4 generation and 5 generation mix together, and their mass matrix is $4\times4$.
With the base $(\tilde{e}_4,\tilde{e}_4^{c*},\tilde{e}_5,\tilde{e}_5^{c*})$, we show the elements of exotic slepton
mass matrix $\mathcal{M}^2_{\tilde{E}}$ in the following form.
\begin{eqnarray}
&&\mathcal{M}^2_{\tilde{E}}(\tilde{e}_5^{c*}\tilde{e}_5^{c})=
\lambda_L^2\frac{\bar{\upsilon}_{NL}^2}{2}+\frac{\upsilon_u^2}{2}|Y_{e_5}|^2+M^2_{\tilde{L}_5}
-\frac{g_1^2-g_2^2}{8}(\upsilon_d^2-\upsilon_u^2)-g_L^2(3+L_4)V_L^2,
\nonumber\\&&\mathcal{M}^2_{\tilde{E}}(\tilde{e}_5^{*}\tilde{e}_5)=\lambda_E^2\frac{\upsilon_{NL}^2}{2}
 +\frac{\upsilon_u^2}{2}|Y_{e_5}|^2+M^2_{\tilde{e}_5}+\frac{g_1^2}{4}(\upsilon_d^2-\upsilon_u^2)
 +g_L^2(3+L_4)V_L^2,
\nonumber\\&&\mathcal{M}^2_{\tilde{E}}(\tilde{e}_4^{*}\tilde{e}_4)=\lambda_L^2\frac{\bar{\upsilon}_{NL}^2}{2}
+\frac{g_1^2-g_2^2}{8}(\upsilon_d^2-\upsilon_u^2)+\frac{\upsilon_d^2}{2}|Y_{e_4}|^2+M^2_{\tilde{L}_4}
+g_L^2L_4V_L^2,
\nonumber\\&&\mathcal{M}^2_{\tilde{E}}(\tilde{e}_4^{c*}\tilde{e}_4^{c})=
\lambda_E^2\frac{\upsilon_{NL}^2}{2}-\frac{g_1^2}{4}(\upsilon_d^2-\upsilon_u^2)+\frac{\upsilon_d^2}{2}|Y_{e_4}|^2+M^2_{\tilde{e}_4}
-g_L^2L_4V_L^2,
\nonumber\\&&\mathcal{M}^2_{\tilde{E}}(\tilde{e}_4^{*}\tilde{e}_5)
=\upsilon_dY_{e_4}^*\lambda_E\frac{\upsilon_{NL}}{2}+\lambda_LY_{e_5}\frac{\bar{\upsilon}_{NL}v_u}{2},
~~~\mathcal{M}^2_{\tilde{E}}(\tilde{e}_5\tilde{e}_5^{c})=\mu^*\frac{\upsilon_d}{\sqrt{2}}Y_{e_5}+A_{e_5}Y_{e_5}\frac{\upsilon_u}{\sqrt{2}},
\nonumber\\&&\mathcal{M}^2_{\tilde{E}}(\tilde{e}_4^{c}\tilde{e}_5)=\mu_{NL}^*\lambda_E
\frac{\bar{\upsilon}_{NL}}{\sqrt{2}}-A_{LE}\lambda_E\frac{\upsilon_{NL}}{\sqrt{2}},
~~\mathcal{M}^2_{\tilde{E}}(\tilde{e}_4\tilde{e}_5^{c})=-\mu_{NL}^*\frac{\upsilon_{NL}}{\sqrt{2}}\lambda_L+A_{LL}\lambda_L\frac{\bar{\upsilon}_{NL}}{\sqrt{2}},
\nonumber\\&&\mathcal{M}^2_{\tilde{E}}(\tilde{e}_4\tilde{e}_4^{c})=\mu^*\frac{\upsilon_u}{\sqrt{2}}Y_{e_4}+A_{e_4}Y_{e_4}\frac{\upsilon_d}{\sqrt{2}},
~~~\mathcal{M}^2_{\tilde{E}}(\tilde{e}_5^{c}\tilde{e}_4^{c*})
=-Y_{e_5}\lambda_E\frac{\upsilon_u\upsilon_{NL}}{2}-\lambda_LY_{e_4}^*\frac{\bar{\upsilon}_{NL}v_d}{2}. \label{SE45}
\end{eqnarray}
In Eq.(\ref{SE45}), the non-zero terms $\mathcal{M}^2_{\tilde{E}}(\tilde{e}_4\tilde{e}_5^{c}), \mathcal{M}^2_{\tilde{E}}(\tilde{e}_4^{*}\tilde{e}_5),
 \mathcal{M}^2_{\tilde{E}}(\tilde{e}_5^{c}\tilde{e}_4^{c*})$ and $\mathcal{M}^2_{\tilde{E}}(\tilde{e}_4^{c}\tilde{e}_5)$ are the reason for
 the exotic slepton mixing of generations 4 and 5. These mixing terms all include the parameters $\upsilon_{NL}$ and $\bar{\upsilon}_{NL}$.
It shows that this mixing is caused basically by the added Higgs superfields $\Phi_{NL}$ and $\varphi_{NL}$.
Using the matrix $Z_{\tilde{E}}$, we obtain mass eigenstates with the formula
$Z^{\dag}_{\tilde{E}}\mathcal{M}^2_{\tilde{E}} Z_{\tilde{E}}=diag(m^2_{\tilde{E}^1},m^2_{\tilde{E}^2},m^2_{\tilde{E}^3},m^2_{\tilde{E}^4})$.

In the same way, the exotic sneutrino mass squared matrix is also obtained
\begin{eqnarray}
\nonumber\\&&\mathcal{M}^2_{\tilde{N}}(\tilde{\nu}_5^{c*}\tilde{\nu}_5^{c})=\lambda_L^2\frac{\bar{\upsilon}_{NL}^2}{2}
-\frac{g_1^2+g_2^2}{8}(\upsilon_d^2-\upsilon_u^2) +\frac{\upsilon_d^2}{2}|Y_{\nu_5}|^2+M^2_{\tilde{L}_5}
-g_L^2(3+L_4)V_L^2,
   \nonumber\\&&\mathcal{M}^2_{\tilde{N}}(\tilde{\nu}_4^{*}\tilde{\nu}_4)
   =\lambda_L^2\frac{\bar{\upsilon}_{NL}^2}{2}+\frac{g_1^2+g_2^2}{8}(\upsilon_d^2-\upsilon_u^2)
    +\frac{\upsilon_u^2}{2}|Y_{\nu_4}|^2+M^2_{\tilde{L}_4}
  +g_L^2L_4V_L^2,
\nonumber\\&&\mathcal{M}^2_{\tilde{N}}(\tilde{\nu}_5^{*}\tilde{\nu}_5)=
\lambda_{NL}^2\frac{\upsilon_{NL}^2}{2}+g_L^2(3+L_4)V_L^2
   +\frac{\upsilon_d^2}{2}|Y_{\nu_5}|^2+M^2_{\tilde{\nu}_5},
\nonumber\\&&\mathcal{M}^2_{\tilde{N}}(\tilde{\nu}_4^{c*}\tilde{\nu}_4^{c})=
\lambda_{NL}^2\frac{\upsilon_{NL}^2}{2}-g_L^2L_4V_L^2
   +\frac{\upsilon_u^2}{2}|Y_{\nu_4}|^2+M^2_{\tilde{\nu}_4},
\nonumber\\&&\mathcal{M}^2_{\tilde{N}}(\tilde{\nu}_5^{c}\tilde{\nu}_4^{c*})=
\lambda_{NL}Y_{\nu_5}\frac{\upsilon_{NL}\upsilon_d}{2}-\lambda_LY_{\nu_4}^*\frac{\bar{\upsilon}_{NL}\upsilon_u}{2},
~~~\mathcal{M}^2_{\tilde{N}}(\tilde{\nu}_5\tilde{\nu}_5^{c})= \mu^*\frac{\upsilon_u}{\sqrt{2}}Y_{\nu_5}+A_{\nu_5}Y_{\nu_5}\frac{\upsilon_d}{\sqrt{2}},
\nonumber\\&&
\mathcal{M}^2_{\tilde{N}}(\tilde{\nu}_4^{c}\tilde{\nu}_5)
=\mu_{NL}^*\lambda_{NL}\frac{\bar{\upsilon}_{NL}}{\sqrt{2}}-A_{LN}\lambda_N\frac{\upsilon_{NL}}{\sqrt{2}},
~~~\mathcal{M}^2_{\tilde{N}}(\tilde{\nu}_4\tilde{\nu}_5^{c})=\mu_{NL}^*\frac{\upsilon_{NL}}{\sqrt{2}}
\lambda_L-A_{LL}\lambda_L\frac{\bar{\upsilon}_{NL}}{\sqrt{2}},
\nonumber\\&&\mathcal{M}^2_{\tilde{N}}(\tilde{\nu}_4^{*}\tilde{\nu}_5)
=\lambda_LY_{\nu_5}\frac{\bar{\upsilon}_{NL}\upsilon_d}{2}-\frac{\upsilon_u\upsilon_{NL}}{2}\lambda_{NL}Y_{\nu_4}^*,~~~
\mathcal{M}^2_{\tilde{N}}(\tilde{\nu}_4\tilde{\nu}_4^{c})=\mu^*\frac{\upsilon_d}{\sqrt{2}}Y_{\nu_4}+A_{\nu_4}Y_{\nu_4}\frac{\upsilon_u}{\sqrt{2}}.
\end{eqnarray}
For the exotic sneutrino, the mixing of generations 4 and 5 is similar as that of exotic slepton.
In the base $(\tilde{\nu}_4,\tilde{\nu}_4^{c*},\tilde{\nu}_5,\tilde{\nu}_5^{c*})$, we get the mass squared matrix of the exotic sneutrino,
and obtain the mass eigenstates by the matrix $Z_{\tilde{N}}$ through the formula
$Z^{\dag}_{\tilde{N}}\mathcal{M}^2_{\tilde{N}} Z_{\tilde{N}}=diag(m^2_{\tilde{N}^1},m^2_{\tilde{N}^2},m^2_{\tilde{N}^3},m^2_{\tilde{N}^4})$.

\subsection{The lepton neutralino mass matrix in EBLMSSM}
In EBLMSSM, the superfields ($\Phi_{L},\varphi_{L},\Phi_{NL},\varphi_{NL}$) have their SUSY
 superpartners $(\psi_{\Phi_L},\psi_{\varphi_L},\psi_{\Phi_{NL}},\psi_{\varphi_{NL}})$. They mix with $\lambda_L$, which is the superpartner
of the new lepton type gauge boson $Z^\mu_L$. Therefore, we deduce their mass matrix in the base
 $(i\lambda_L,\psi_{\Phi_L},\psi_{\varphi_L},\psi_{\Phi_{NL}},\psi_{\varphi_{NL}})$
\begin{equation}
\mathcal{M}_L=\left(     \begin{array}{ccccc}
  2M_L &2\upsilon_Lg_L &-2\bar{\upsilon}_Lg_L&3\upsilon_{NL}g_L &-3\bar{\upsilon}_{NL}g_L\\
   2\upsilon_Lg_L & 0 &-\mu_L& 0 & 0\\
   -2\bar{\upsilon}_Lg_L&-\mu_L &0& 0 & 0\\
   3\upsilon_{NL}g_L & 0 & 0 & 0 & -\mu_{NL}\\
   -3\bar{\upsilon}_{NL}g_L& 0&0&-\mu_{NL}&0
    \end{array}\right).
\end{equation}
The lepton neutralino mass egeinstates are four-component spinors $X^0_{L_i}=(K_{L_i}^0,\bar{K}_{L_i}^0)^T$,
and their mass matrix is diagonalized by the rotation matrix $Z_{NL}$. The relations for the components are
\begin{eqnarray}
&&i\lambda_L=Z_{NL}^{1i}K_{L_i}^0
,~~~\psi_{\Phi_L}=Z_{NL}^{2i}K_{L_i}^0
,~~~\psi_{\varphi_L}=Z_{NL}^{3i}K_{L_i}^0,
\nonumber\\&&\psi_{\Phi_{NL}}=Z_{NL}^{4i}K_{L_i}^0
,~~~~~\psi_{\varphi_{NL}}=Z_{NL}^{5i}K_{L_i}^0.
\end{eqnarray}
In BLMSSM, there are no $\psi_{\Phi_{NL}},\psi_{\varphi_{NL}}$, and the base of lepton neutralino is
$(i\lambda_L,\psi_{\Phi_L},\psi_{\varphi_L})$, whose mass matrix is $3\times3$. EBLMSSM extends this matrix to $5\times5$
including the BLMSSM results.

\subsection{The Higgs superfields and $Y$ in EBLMSSM}
The superfields $\Phi_{L},\varphi_{L},\Phi_{NL},\varphi_{NL}$ mix together and form $4\times4$ mass squared matrix, which is larger than the
corresponding $2\times2$ mass matrix in the BLMSSM.
Diagonalizing the mass squared matrix, four CP even exotic Higgs are obtained.
\begin{eqnarray}
&&\mathcal{M}^2_{\phi}(\Phi_L^0\Phi_L^0)=\frac{1}{2}g_L^2\Big(6\upsilon_L^2-2\bar{\upsilon}_L^2+3(\upsilon_{NL}^2-\bar{\upsilon}_{NL}^2)\Big)
+\frac{1}{2}\mu_L^2+\frac{1}{2}m_{\Phi_L}^2,
\nonumber\\&&\mathcal{M}^2_{\phi}(\varphi_L^0\varphi_L^0)=
\frac{1}{2}g_L^2\Big(6\bar{\upsilon}_L^2-2\upsilon_L^2+3(\bar{\upsilon}_{NL}^2-\upsilon_{NL}^2)\Big)+\frac{1}{2}\mu_L^2
+\frac{1}{2}m_{\varphi_L}^2,
\nonumber\\&&\mathcal{M}^2_{\phi}(\Phi_{NL}^0\Phi_{NL}^0)
=\frac{1}{2}g_L^2\Big(\frac{27}{2}\upsilon_{NL}^2-\frac{9}{2}\bar{\upsilon}_{NL}^2+3(\upsilon_L^2-\bar{\upsilon}_L^2)\Big)
+\frac{1}{2}\mu_{NL}^2+\frac{1}{2}m_{\Phi_{NL}}^2,
\nonumber\\&&\mathcal{M}^2_{\phi}(\varphi_{NL}^0\varphi_{NL}^0)
=\frac{1}{2}g_L^2\Big(\frac{27}{2}\bar{\upsilon}_{NL}^2-\frac{9}{2}\upsilon_{NL}^2+3(\bar{\upsilon}_L^2-\upsilon_L^2)\Big)
+\frac{1}{2}\mu_{NL}^2+\frac{1}{2}m_{\varphi_{NL}}^2,
\nonumber\\&&\mathcal{M}^2_{\phi}(\Phi_L^0\varphi_L^0)=-4g_L^2\upsilon_L\bar{\upsilon}_L-\frac{B_L\mu_L}{2},~~~~~~~~~~~~
\mathcal{M}^2_{\phi}(\Phi_L^0\Phi_{NL}^0)=6g_L^2\upsilon_L\upsilon_{NL},
\nonumber\\&&\mathcal{M}^2_{\phi}(\Phi_{NL}^0\varphi_{NL}^0)=-9g_L^2\upsilon_{NL}\bar{\upsilon}_{NL}-\frac{B_{NL}\mu_{NL}}{2},
~~~\mathcal{M}^2_{\phi}(\varphi_L^0\varphi_{NL}^0)=6g_L^2\bar{\upsilon}_L\bar{\upsilon}_{NL},
\nonumber\\&&\mathcal{M}^2_{\phi}(\varphi_L^0\Phi_{NL}^0)=-6g_L^2\bar{\upsilon}_L\upsilon_{NL},~~~~~~~~~~~~~~~~~~~~~
\mathcal{M}^2_{\phi}(\Phi_L^0\varphi_{NL}^0)=-6g_L^2\upsilon_L\bar{\upsilon}_{NL}.\label{Phimass}
\end{eqnarray}
We use $Z_{\tilde{\phi}_L}$ to diagonalize the mass squared matrix in Eq.(\ref{Phimass}), and the relation between mass eigenstates and the comments are
\begin{eqnarray}
&&\Phi_L^0=Z_{\tilde{\phi}_L}^{1i}H_{L_i}^0
,~~~\varphi_L^0=Z_{\tilde{\phi}_L}^{2i}H_{L_i}^0, ~~~
\Phi_{NL}^0=Z_{\tilde{\phi}_{L}}^{3i}H_{L_i}^0
,~~~\varphi_{NL}^0=Z_{\tilde{\phi}_{L}}^{4i}H_{L_i}^0.
\end{eqnarray}

In EBLMSSM, the conditions for the exotic CP odd Higgs $P_L^0, \bar{P}_L^0$ are same as those in BLMSSM, and
 they do not mix with the added exotic CP odd Higgs $P_{NL}^0, \bar{P}_{NL}^0$. Here, we show
 the mass squared matrix for the added exotic CP odd Higgs $P_{NL}^0, \bar{P}_{NL}^0$.
\begin{eqnarray}
&&\mathcal{M}^2_{p}(P_{NL}^0P_{NL}^0)=\frac{1}{2}g_L^2\Big(\frac{9}{2}\upsilon_{NL}^2
-\frac{9}{2}\bar{\upsilon}_{NL}^2+3(\upsilon_L^2-\bar{\upsilon}_L^2)\Big)+\frac{1}{2}\mu_{NL}^2+\frac{1}{2}m_{\Phi_{NL}}^2,
\nonumber\\&&\mathcal{M}^2_{p}(\bar{P}_{NL}^0\bar{P}_{NL}^0)=\frac{1}{2}g_L^2
\Big(\frac{9}{2}\bar{\upsilon}_{NL}^2-\frac{9}{2}\upsilon_{NL}^2+3(\bar{\upsilon}_L^2
-\upsilon_L^2)\Big)+\frac{1}{2}\mu_{NL}^2+\frac{1}{2}m_{\varphi_{NL}}^2,
\nonumber\\&&\mathcal{M}^2_{p}(P_{NL}^0\bar{P}_{NL}^0)=\frac{B_{NL}\mu_{NL}}{2}.
\end{eqnarray}

The scalar superfields $Y$ and $Y'$ mix, and their mass squared matrix is deduced here. This condition is similar as that of
$X$ and $X'$, then the lightest mass egeinstate of  $Y$ and $Y'$ can be a candidate of the dark matter.
 With $S_{Y}=g_{L}^2(2+L_{4})V_L^2$,
 the concrete form for the mass squared matrix is shown here. To obtain mass eigenstates, the matrix $Z_Y$ is
 used through the following formula, with the supposition $m_{{Y_1}}^2<m_{{Y_2}}^2$.
  \begin{eqnarray}
Z^{\dag}_{Y}\left(     \begin{array}{cc}
  |\mu_{Y}|^2+S_{Y} &-\mu_{Y}B_{Y} \\
    -\mu^*_{Y}B^*_{Y} & |\mu_{Y}|^2-S_{Y}\\
    \end{array}\right)  Z_{Y}=\left(     \begin{array}{cc}
 m_{{Y_1}}^2 &0 \\
    0 & m_{{Y_2}}^2\\
    \end{array}\right),
   ~~~~~\left(     \begin{array}{c}
  Y_{1} \\  Y_{2}\\
    \end{array}\right) =Z_{Y}^{\dag}\left( \begin{array}{c}
  Y \\  Y'^*\\
    \end{array}\right).\label{YY'}
   \end{eqnarray}
The superpartners of $Y$ and $Y'$ form four-component Dirac spinors, and the mass term for superfields $\tilde{Y}$ is shown as
\begin{eqnarray}
  &&-\mathcal{L}^{mass}_{\tilde{Y}}=\mu_Y\bar{\tilde{Y}}\tilde{Y}
  ,~~~~~~~~~~~~~~~~\tilde{Y} =\left( \begin{array}{c}
  \psi_{Y'} \\  \bar{\psi}_{Y}\\
    \end{array}\right).
\end{eqnarray}
The spinor $\tilde{Y}$ and the mixing of superfields $Y,Y'$ are all new terms beyond BLMSSM, that add abundant contents to lepton physics and dark matter
physics.

\subsection{Some couplings with $h^0$ in EBLMSSM}
In EBLMSSM, the exotic slepton(sneutrino) of generations 4 and 5 mix.
So the couplings with exotic slepton(sneutrino) are different from the corresponding results
in BLMSSM. We deduce the couplings of $h^0$ and exotic sleptons
\begin{eqnarray}
&&\sum_{i,j=1}^4\tilde{E}^{i*}\tilde{E}^{j}h^0\Big[\Big(e^2\upsilon\sin\beta\frac{1-4s_W^2}{4s_W^2c_W^2}(Z_{\tilde{E}}^{4i*}Z_{\tilde{E}}^{4j}
-Z_{\tilde{E}}^{1i*}Z_{\tilde{E}}^{1j})
    -\frac{\mu^*}{\sqrt{2}}Y_{e_4}Z_{\tilde{E}}^{2i*}Z_{\tilde{E}}^{1j}\nonumber\\&&
    -\upsilon\sin\beta|Y_{e_5}|^2\delta_{ij}-\frac{A_{E_5}}{\sqrt{2}}Z_{\tilde{E}}^{4i*}Z_{\tilde{E}}^{3j}
    +\frac{1}{2}\lambda_LY_{e_5}Z_{\tilde{E}}^{3j}Z_{\tilde{E}}^{3i*}\bar{\upsilon}_{NL}
-\frac{1}{2}Y_{e_5}^*Z_{\tilde{E}}^{4j}\lambda_EZ_{\tilde{E}}^{2i*}\upsilon_{NL}\Big)\cos\alpha\nonumber\\&&-
    \Big(e^2\upsilon\cos\beta\frac{1-4s_W^2}{4s_W^2c_W^2}(Z_{\tilde{E}}^{1i*}Z_{\tilde{E}}^{1j}-Z_{\tilde{E}}^{4i*}Z_{\tilde{E}}^{4j})
  -\upsilon\cos\beta|Y_{e_4}|^2\delta_{ij}-\frac{A_{E_4}}{\sqrt{2}}Z_{\tilde{E}}^{2i*}Z_{\tilde{E}}^{1j}\nonumber\\&&
    -\frac{\mu^*}{\sqrt{2}}Y_{e_5}Z_{\tilde{E}}^{4i*}Z_{\tilde{E}}^{3j}-\frac{1}{2}Y_{e_4}^*Z_{\tilde{E}}^{2j}\lambda_LZ_{\tilde{E}}^{4i*}\bar{\upsilon}_{NL}
+\frac{1}{2}Z_{\tilde{E}}^{1i*}Y_{e_4}^*\lambda_EZ_{\tilde{E}}^{3j}\upsilon_{NL}
    \Big)\sin\alpha\Big].\label{hEECP}
\end{eqnarray}
In Eq.(\ref{hEECP}), different from BLMSSM, there are new terms  $(\frac{1}{2}\lambda_LY_{e_5}Z_{\tilde{E}}^{3j}Z_{\tilde{E}}^{3i*}\bar{\upsilon}_{NL}
-\frac{1}{2}Y_{e_5}^*Z_{\tilde{E}}^{4j}\lambda_EZ_{\tilde{E}}^{2i*}\upsilon_{NL})\cos\alpha-
(\frac{1}{2}Z_{\tilde{E}}^{1i*}Y_{e_4}^*\lambda_EZ_{\tilde{E}}^{3j}\upsilon_{NL}
-\frac{1}{2}Y_{e_4}^*Z_{\tilde{E}}^{2j}\lambda_LZ_{\tilde{E}}^{4i*}\bar{\upsilon}_{NL})\sin\alpha$ besides the mixing of generations 4 and 5 slepton.
Obviously, these new terms include
$\upsilon_{NL}$ and $\bar{\upsilon}_{NL}$, which are the VEVs of added Higgs superfields $\Phi_{NL}$ and $\varphi_{NL}$.
In the same way, the couplings of $h^0$ and exotic sneutrinos are also calculated
\begin{eqnarray}
&&\sum_{i,j=1}^4\tilde{N}^{i*}\tilde{N}^{j}h^0\Big[\Big(\frac{e^2}{4s_W^2c_W^2}\upsilon\sin\beta
(Z_{\tilde{N}}^{1i*}Z_{\tilde{N}}^{1j}-Z_{\tilde{N}}^{4i*}Z_{\tilde{N}}^{4j})
  -\frac{1}{2}Z_{\tilde{N}}^{1i*}Y_{\nu_4}^*\lambda_{NL}Z_{\tilde{N}}^{3i}\upsilon_{NL}
  \nonumber\\&&-\upsilon\sin\beta|Y_{\nu_4}|^2\delta_{ij}-\frac{A_{N_4}}{\sqrt{2}}Z_{\tilde{N}}^{2i*}Z_{\tilde{N}}^{1j}-\frac{\mu^*}{\sqrt{2}}Y_{\nu_5}
  Z_{\tilde{N}}^{4i*}Z_{\tilde{N}}^{3j}-\frac{1}{2}Y_{\nu_4}^*Z_{\tilde{N}}^{2j}\lambda_LZ_{\tilde{N}}^{4i*}\bar{\upsilon}_{NL}
\Big)\cos\alpha\nonumber\\&&-\Big(\frac{e^2}{4s_W^2c_W^2}\upsilon\cos\beta [Z_{\tilde{N}}^{4i*}Z_{\tilde{N}}^{4j}
-Z_{\tilde{N}}^{1i*}Z_{\tilde{N}}^{1j}]-\frac{\mu^*}{\sqrt{2}}Y_{\nu_4}
  Z_{\tilde{N}}^{2i*}Z_{\tilde{N}}^{1j}-\upsilon\cos\beta|Y_{\nu_5}|^2\delta_{ij}\nonumber\\&&
  -\frac{A_{N_5}}{\sqrt{2}}Z_{\tilde{N}}^{4i*}Z_{\tilde{N}}^{3j}+\frac{1}{2}Y_{\nu_5}Z_{\tilde{N}}^{3j}
  \lambda_LZ_{\tilde{N}}^{1i*}\bar{\upsilon}_{NL}
+\frac{1}{2}Y_{\nu_5}Z_{\tilde{N}}^{4i*}\lambda_{N^c}Z_{\tilde{N}}^{2j}\upsilon_{NL}
\Big)\sin\alpha\Big].
\end{eqnarray}
In this coupling, the new terms beyond BLMSSM are
$-(\frac{1}{2}Z_{\tilde{N}}^{1i*}Y_{\nu_4}^*\lambda_{NL}Z_{\tilde{N}}^{3i}\upsilon_{NL}
  +\frac{1}{2}Y_{\nu_4}^*Z_{\tilde{N}}^{2j}\lambda_LZ_{\tilde{N}}^{4i*}\bar{\upsilon}_{NL})\cos\alpha
 -(\frac{1}{2}Y_{\nu_5}Z_{\tilde{N}}^{3j}
  \lambda_LZ_{\tilde{N}}^{1i*}\bar{\upsilon}_{NL}
+\frac{1}{2}Y_{\nu_5}Z_{\tilde{N}}^{4i*}\lambda_{N^c}Z_{\tilde{N}}^{2j}\upsilon_{NL})\sin\alpha.$

The $h^0-\tilde{L}-\tilde{L}$ coupling has the same form as that in BLMSSM.
While, the $h^0-\tilde{\nu}-\tilde{\nu}$ coupling gets corrected
 terms, but these terms are suppressed by the tiny neutrino Yukawa
 coupling $Y_\nu$.
\begin{eqnarray}
&&\sum_{i,j=1}^6\tilde{\nu}^{i*}\tilde{\nu}^jh^0\Big[\sin\alpha\frac{\mu^*}{\sqrt{2}}Y_{\nu}^*Z_{\tilde{\nu}}^{Ii*}Z_{\tilde{\nu}}^{(I+3)j}
-\frac{e^2}{4s_W^2c_W^2}B_R^2Z_{\tilde{\nu}}^{Ii*}Z_{\tilde{\nu}}^{Ij}\nonumber\\&&+\cos\alpha\Big(
(\lambda_{N^c}\bar{\upsilon}_L
-\frac{A_N}{\sqrt{2}})Y_{\nu}^*Z_{\tilde{\nu}}^{Ii*}Z_{\tilde{\nu}}^{(I+3)j}-\upsilon\sin\beta|Y_{\nu}|^2\delta_{ij}
\Big)\Big].
\end{eqnarray}
Here, $s_W(c_W)$ denotes $\sin\theta_W(\cos\theta_W)$, with $\theta_W$ representing the weak-mixing angle.
The concrete form of $B_R^2$ is in Ref.\cite{MSSM}.

\subsection{The couplings with $Y$}
For the dark matter candidate $Y_1$, the necessary tree level couplings are deduced in EBLMSSM.
We show the couplings (lepton-exotic lepton-$Y$) and (neutrino-exotic neutrino-$Y$)
\begin{eqnarray}
&&\mathcal{L}=\sum_{i,j=1}^2\bar{e}^I\Big(\lambda_4W_L^{1i}Z_Y^{1j*}P_R -\lambda_6U_L^{2i}Z_Y^{2j*}P_L\Big)L'_{i+3}Y_j^*\nonumber\\&&
-\sum_{\alpha=1}^6\sum_{i,j=1}^2\bar{X}_{N_\alpha}^0\Big(\lambda_4Z_{N_{\nu}}^{I\alpha*}W_N^{1i}Z_Y^{1j*}
P_R+\lambda_5Z_{N_{\nu}}^{(I+3)\alpha}U_N^{2i}Z_Y^{2j*}P_L\Big)
N'_{i+3}Y_j^*+h.c.\label{TCLY}
\end{eqnarray}
The new gauge boson $Z_L$ couples with leptons, neutrinos and $Y$, whose concrete forms are
\begin{eqnarray}
&&\mathcal{L}=-\sum_{I=1}^3g_LZ^\mu_L\bar{e}^I\gamma_\mu e^I-\sum_{i,j=1}^2g_L(2+L_4)Z^\mu_LY_i^*i\partial_\mu Y_j
\nonumber\\&&-\sum_{I=1}^3\sum_{\alpha,\beta=1}^6g_LZ^\mu_L\bar{\chi}^0_{N_\alpha}(Z_{N_\nu}^{I\alpha*}Z_{N_\nu}^{I\beta}\gamma_\mu P_L
+Z_{N_\nu}^{(I+3)\alpha*}Z_{N_\nu}^{(I+3)\beta}\gamma_\mu  P_R)\chi^0_{N_\beta}+h.c.\label{TCZL}
\end{eqnarray}
$\varphi_L$ gives masses to the light neutrinos trough the see-saw mechanism and $\Phi_L,\varphi_L, \Phi_{NL},\varphi_{NL}$ mix together
producing lepton Higgs $H^0_{L}$. Then the couplings of $H^0_L YY^*$ and $\bar{\chi}^0_{N}\chi^0_{N}H^0_{L}$ are needed
\begin{eqnarray}
&&\mathcal{L}=\sum_{i,j=1}^2\sum_{k=1}^4g_L^2(2+L_4)(Z_Y^{1i*}Z_Y^{1j}-Z_Y^{2i*}Z_Y^{2j})\nonumber\\&&\times
\Big(v_LZ^{1k}_{\tilde{\phi}_L}-\bar{v}_LZ^{2k}_{\tilde{\phi}_L}
+\frac{3}{2}v_{NL}Z^{3k}_{\tilde{\phi}_{NL}}-\frac{3}{2}\bar{v}_{NL}Z^{4k}_{\tilde{\phi}_{NL}}\Big)H^0_{L_k}Y_i^*Y_j.
\nonumber\\&&-\sum_{k=1}^4\sum_{\alpha,\beta=1}^6\lambda_{N^c}Z_{N_\nu}^{(I+3)\alpha}
Z_{N_\nu}^{(I+3)\beta}Z_{\phi_L}^{2k}\bar{\chi}^0_{N_\alpha}
P_L\chi^0_{N_\beta}H^0_{L_k}+h.c.\label{TCHL}
\end{eqnarray}
\section{The  mass of $h^0$}

Similar as BLMSSM,  in EBLMSSM the mass squared matrix for the
neutral CP even Higgs are studied, and in the basis $(H_d^0,\;H_u^0)$ it is written as
\begin{eqnarray}
&&{\cal M}^2_{even}=\left(\begin{array}{ll}M_{11}^2+\Delta_{11}&M_{12}^2+\Delta_{12}\\
M_{12}^2+\Delta_{12}&M_{22}^2+\Delta_{22}\end{array}\right)\;,
\label{MCPEHiggs}
\end{eqnarray}
where
$M_{11}^2,M_{12}^2,M_{22}^2$ are the tree level results, whose concrete forms can be found in Ref.\cite{TFBL}

\begin{eqnarray}
&&\Delta_{11}=\Delta_{11}^{MSSM}+\Delta_{11}^{B}+\Delta_{11}^{L}\;,
\nonumber\\
&&\Delta_{12}=\Delta_{12}^{MSSM}+\Delta_{12}^{B}+\Delta_{12}^{L}\;,
\nonumber\\
&&\Delta_{22}=\Delta_{22}^{MSSM}+\Delta_{22}^{B}+\Delta_{22}^{L}\;.
\label{M-CPE2}
\end{eqnarray}
The MSSM contributions are represented by $\Delta_{11}^{MSSM}$, $\Delta_{12}^{MSSM}$ and $\Delta_{22}^{MSSM}$.
The exotic quark(squark) contributions denoted by $\Delta_{11}^{B},\Delta_{12}^{B}$ and $\Delta_{22}^{B}$ are the same as those in BLMSSM\cite{TFBL}.
However, the corrections $\Delta_{11}^{L},\Delta_{12}^{L}$ and $\Delta_{22}^{L}$ from exotic lepton(slepton) are different from those in BLMSSM,
because the mass squared matrices of exotic slepton and exotic sneutrino are both $4\times4$ and  they relate with $\upsilon_{NL}$ and $\bar{\upsilon}_{NL}$. Furthermore,
the exotic leptons and exotic neutrinos are heavier than those in BLMSSM, due to the introduction of $\Phi_{NL}$ and $\varphi_{NL}$.

\begin{eqnarray}
&&\Delta_{11}^{L}={G_{F}Y_{\nu_4}^4\upsilon^4\over4\sqrt{2}\pi^2\sin^2\beta}\cdot
{\mu^2(A_{{\nu_4}}-\mu\cot\beta)^2\over(m_{{\tilde{N}^1}}^2-m_{{\tilde{N}^2}}^2)^2}
g(m_{{\tilde{N}^1}},m_{{\tilde{N}^2}})
+{G_{F}Y_{{\nu_5}}^4\upsilon^4\over4\sqrt{2}\pi^2\cos^2\beta}\Big\{\ln{m_{{\tilde{N}^3}}m_{{\tilde{N}^4}}
\over m_{{\nu_5}}^2}\nonumber\\&&\hspace{1.2cm}+{A_{{\nu_5}}(A_{{\nu_5}}-\mu\tan\beta)\over m_{{\tilde{N}^3}}^2-m_{{\tilde{N}^4}}^2}
\ln{m_{{\tilde{N}^3}}^2\over m_{{\tilde{N}^4}}^2}
+{A_{{\nu_5}}^2(A_{{\nu_5}}-\mu\tan\beta)^2\over(m_{{\tilde{N}^3}}^2-m_{{\tilde{N}^4}}^2)^2}
g(m_{{\tilde{N}^3}}, m_{{\tilde{N}^4}})\Big\}
\nonumber\\&&\hspace{1.2cm}
+{G_{F}Y_{{e_4}}^4\upsilon^4\over4\sqrt{2}\pi^2\cos^2\beta}\Big\{{A_{{e_4}}(A_{{e_4}}-\mu\tan\beta)\over m_{{\tilde{E}^1}}^2-m_{{\tilde{E}^2}}^2}
\ln{m_{{\tilde{E}^1}}^2\over m_{{\tilde{E}^2}}^2}
+{A_{{e_4}}^2(A_{{e_4}}-\mu\tan\beta)^2\over(m_{{\tilde{E}^1}}^2-m_{{\tilde{E}^2}}^2)^2}
g(m_{{\tilde{E}^1}}, m_{{\tilde{E}^2}})
\nonumber\\&&\hspace{1.2cm}+\ln{m_{{\tilde{E}^1}}m_{{\tilde{E}^2}}
\over m_{{e_4}}^2}\Big\}
+{G_{F}Y_{{e_5}}^4\upsilon^4\over4\sqrt{2}\pi^2\sin^2\beta}\cdot
{\mu^2(A_{{e_5}}-\mu\cot\beta)^2\over(m_{{\tilde{E}^3}}^2-m_{{\tilde{E}^4}}^2)^2}
g(m_{{\tilde{E}^3}},m_{{\tilde{E}^4}})
\;,\nonumber\\
&&\Delta_{12}^{L}={G_{F}Y_{{\nu_4}}^4\upsilon^4\over4\sqrt{2}\pi^2\sin^2\beta}\cdot
{\mu(\mu\cot\beta-A_{{\nu_4}})\over m_{{\tilde{N}^1}}^2-m_{{\tilde{N}^2}}^2}
\Big\{\ln{m_{{\tilde{N}^1}}\over m_{{\tilde{N}^2}}}+{A_{{\nu_4}}(A_{{\nu_4}}-\mu\cot\beta)
\over m_{{\tilde{N}^1}}^2-m_{{\tilde{N}^2}}^2}g(m_{{\tilde{N}^1}},m_{{\tilde{N}^2}})\Big\}
\nonumber\\
&&\hspace{1.2cm}
+{G_{F}Y_{{e_4}}^4\upsilon^4\over4\sqrt{2}\pi^2\cos^2\beta}\cdot
{\mu(\mu\tan\beta-A_{{e_4}})\over m_{{\tilde{E}^1}}^2-m_{{\tilde{E}^2}}^2}
\Big\{\ln{m_{{\tilde{E}^1}}\over m_{{\tilde{E}^2}}}+{A_{{e_4}}(A_{{e_4}}-\mu\tan\beta)
\over m_{{\tilde{E}^1}}^2-m_{{\tilde{E}^2}}^2}g(m_{{\tilde{E}^1}},m_{{\tilde{E}^2}})\Big\}
\nonumber\\
&&\hspace{1.2cm}
+{G_{F}Y_{{\nu_5}}^4\upsilon^4\over4\sqrt{2}\pi^2\cos^2\beta}\cdot
{\mu(\mu\tan\beta-A_{{\nu_5}})\over m_{{\tilde{N}^3}}^2-m_{{\tilde{N}^3}}^2}
\Big\{\ln{m_{{\tilde{N}^3}}\over m_{{\tilde{N}^4}}}+{A_{{\nu_5}}(A_{{\nu_5}}-\mu\tan\beta)
\over m_{{\tilde{N}^3}}^2-m_{{\tilde{N}^4}}^2}g(m_{{\tilde{N}^3}},m_{{\tilde{N}^4}})\Big\}
\nonumber\\
&&\hspace{1.2cm}
+{G_{F}Y_{{e_5}}^4\upsilon^4\over4\sqrt{2}\pi^2\sin^2\beta}\cdot
{\mu(\mu\cot\beta-A_{{e_5}})\over m_{{\tilde{E}^3}}^2-m_{{\tilde{E}^4}}^2}
\Big\{\ln{m_{{\tilde{E}^3}}\over m_{{\tilde{E}^4}}}+{A_{{e_5}}(A_{{e_5}}-\mu\cot\beta)
\over m_{{\tilde{E}^3}}^2-m_{{\tilde{E}^4}}^2}g(m_{{\tilde{E}^3}},m_{{\tilde{E}^4}})\Big\}
\;,\nonumber\\
&&\Delta_{22}^{L}={G_{F}Y_{{\nu_4}}^4\upsilon^4\over4\sqrt{2}\pi^2\sin^2\beta}
\Big\{{A_{{\nu_4}}(A_{{\nu_4}}-\mu\cot\beta)\over m_{{\tilde{N}^1}}^2-m_{{\tilde{N}^2}}^2}
\ln{m_{{\tilde{N}^1}}^2\over m_{{\tilde{N}^2}}^2}
+{A_{{\nu_4}}^2(A_{{\nu_4}}-\mu\cot\beta)^2\over(m_{{\tilde{N}^1}}^2-m_{{\tilde{N}^2}}^2)^2}
g(m_{{\tilde{N}^1}}, m_{{\tilde{N}^2}})
\nonumber\\
&&\hspace{1.2cm}+\ln{m_{{\tilde{N}^1}}m_{{\tilde{N}^2}}
\over m_{{\nu_4}}^2}\Big\}
+{G_{F}Y_{{e_4}}^4\upsilon^4\over4\sqrt{2}\pi^2\cos^2\beta}\cdot
{\mu^2(A_{{e_4}}-\mu\tan\beta)^2\over(m_{{\tilde{E}^1}}^2-m_{{\tilde{E}^2}}^2)^2}
g(m_{{\tilde{E}^1}},m_{{\tilde{E}^2}})
\nonumber\\&&\hspace{1.2cm}
+{G_{F}Y_{{e_5}}^4\upsilon^4\over4\sqrt{2}\pi^2\sin^2\beta}\Big\{{A_{{e_5}}(A_{{e_5}}
-\mu\cot\beta)\over m_{{\tilde{E}^3}}^2-m_{{\tilde{E}^4}}^2}
\ln{m_{{\tilde{E}^3}}^2\over m_{{\tilde{E}^4}}^2}
+{A_{{e_5}}^2(A_{{e_5}}-\mu\cot\beta)^2\over(m_{{\tilde{E}^3}}^2-m_{{\tilde{E}^4}}^2)^2}
g(m_{{\tilde{E}^3}}, m_{{\tilde{E}^4}})\nonumber\\
&&\hspace{1.2cm}+\ln{m_{{\tilde{E}^3}}m_{{\tilde{E}^4}}
\over m_{{e_5}}^2}\Big\}
+{G_{F}Y_{{\nu_5}}^4\upsilon^4\over4\sqrt{2}\pi^2\cos^2\beta}\cdot
{\mu^2(A_{{\nu_5}}-\mu\tan\beta)^2\over(m_{{\tilde{N}^3}}^2-m_{{\tilde{N}^4}}^2)^2}
g(m_{{\tilde{N}^3}},m_{{\tilde{N}^4}})\;.
\label{app4-1}
\end{eqnarray}

\section{the processes $h^0\rightarrow \gamma\gamma, ~h^0\rightarrow VV,~ V=(Z,W)$ and dark matter $Y_1$}
\subsection{$h^0$ decays}
At the LHC, $h^0$ is produced chiefly from the gluon fusion $(gg\rightarrow h^0)$.
The one loop diagrams are the leading order (LO) contributions. The virtual t quark loop is the dominate contribution because
of the large Yukawa coupling.  Therefore, when the couplings of new particles and Higgs are large, they can influence the results obviously.
For $ h^0\rightarrow gg$, the EBLMSSM results are same as those in BLMSSM, and are shown as\cite{htogg,htoxx}
\begin{eqnarray}
&&\Gamma_{{NP}}(h^0\rightarrow gg)={G_{F}\alpha_s^2m_{{h^0}}^3\over64\sqrt{2}\pi^3}
\Big|\sum\limits_{q,q'}g_{{h^0qq}}A_{1/2}(x_q)
+\sum\limits_{\tilde q, \tilde q'}g_{{h^0\tilde{q}\tilde{q}}}{m_{{\rm Z}}^2\over m_{{\tilde q}}^2}A_{0}(x_{{\tilde{q}}})\Big|^2\;,
\label{hgg}
\end{eqnarray}
with $x_a=m_{{h^0}}^2/(4m_a^2)$. Here, $q$ and $q'$ are quark and exotic quark. While, $\tilde{q}$ and $\tilde{q}'$ denote squark and exotic squark.
The concrete expressions for $g_{{h^0qq}},\;g_{{h^0q'q'}},\;g_{{h^0\tilde{q}\tilde{q}}}
,\;g_{{h^0\tilde{q}'\tilde{q}'}}\;(i=1,\;2)$ are in literature \cite{TFBL}.
The functions $A_{1/2}(x)$ and $A_0(x)$ are\cite{htoxx}
\begin{eqnarray}
&&A_{1/2}(x)=2\Big[x+(x-1)g(x)\Big]/x^2,~~~~~~A_0(x)=-(x-g(x))/x^2\;,\nonumber\\
&&g(x)=\left\{\begin{array}{l}\arcsin^2\sqrt{x},\;x\le1\\
-{1\over4}\Big[\ln{1+\sqrt{1-1/x}\over1-\sqrt{1-1/x}}-i\pi\Big]^2,\;x>1\;.\end{array}\right.
\label{g-function}
\end{eqnarray}

The decay $h^0\rightarrow \gamma\gamma$ obtains contributions from loop diagrams, and the leading order contributions
are from the one loop diagrams. In the EBLMSSM, the exotic quark(squark) and exotic lepton (slepton) give
new corrections  to the decay width of $h^0\rightarrow \gamma\gamma$. Different from BLMSSM, the
exotic leptons in EBLMSSM are more heavy and the exotic sleptons of the 4 and 5 generations mix together. These parts should influence the numerical
results of the EBLMSSM theoretical prediction to the process $h^0\rightarrow \gamma\gamma$ to some extent.

The decay width of $h^0\rightarrow \gamma\gamma$ can be expressed as\cite{htopp}
\begin{eqnarray}
&&\Gamma_{{NP}}(h^0\rightarrow\gamma\gamma)={G_{F}\alpha^2m_{{h^0}}^3\over128\sqrt{2}\pi^3}
\Big|\sum\limits_fN_cQ_{f}^2g_{{h^0ff}}A_{1/2}(x_f)+g_{{h^0H^+H^-}}{m_{{\rm W}}^2\over m_{{H^\pm}}^2}A_0(x_{{H^\pm}})
\nonumber\\&&+g_{{h^0WW}}A_1(x_{{\rm W}})
+\sum\limits_{i=1}^2g_{{h^0\chi_i^+\chi_i^-}}{m_{{\rm W}}\over m_{{\chi_i}}}A_{1/2}(x_{{\chi_i}})
+\sum\limits_{\tilde f}N_cQ_{f}^2g_{{h^0\tilde{f}\tilde{f}}}{m_{ Z}^2\over m_{{\tilde f}}^2}
A_{0}(x_{{\tilde{f}}})\Big|^2\;,
\label{hpp}
\end{eqnarray}
where $g_{{h^0WW}}=\sin(\beta-\alpha)$ and $A_1(x)=-\Big[2x^2+3x+3(2x-1)g(x)\Big]/x^2$.

The formulae for $h^0\rightarrow ZZ, WW$ are
\begin{eqnarray}
&&\Gamma(h^0\rightarrow WW)={3e^4m_{{h^0}}\over512\pi^3s_{ W}^4}|g_{h^0WW}|^2
F({m_{_{\rm W}}\over m_{h^0}}),\;\nonumber\\
&&\Gamma(h^0\rightarrow ZZ)={e^4m_{{h^0}}\over2048\pi^3s_{W}^4c_{W}^4}|g_{h^0ZZ}|^2
\Big(7-{40\over3}s_{W}^2+{160\over9}s_{W}^4\Big)F({m_{Z}\over m_{_{h^0}}}),
\end{eqnarray}
with $g_{{h^0ZZ}}=g_{{h^0WW}}$ and $F(x)$ is given out in Ref\cite{htoww,htozz}.
The observed signals for the diphoton and $ZZ,\;WW$ channels are
quantified by the ratios $R_{\gamma\gamma}$ and $R_{VV}, ~V=(Z,W)$, whose current values are
$R_{\gamma\gamma}=1.16\pm0.18$ and $R_{VV}=1.19^{+0.22}_{-0.20}$ \cite{pdg2016}.

\subsection{Dark matter $Y$}

In BLMSSM, there are some dark matter candidates such as: the lightest mass eigenstate of $X X^\prime$ mixing,
$\tilde{X}$ the four-component spinor composed by the super partners of $X$ and $X^\prime$. They are studied in Ref.\cite{darkM}.
In EBLMSSM, the dark matter candidates are more than those in BLMSSM, because the lightest mass eigenstate of $Y  Y^\prime$ mixing and
$\tilde{Y}$ are dark matter candidates. After $U(1)_L$ is broken by $\Phi_L$ and
$\Phi_{NL}$, Z2 symmetry is left, which guarantees the stability of the dark matters.
There are only two elements (1,-1) in Z2 group. This symmetry eliminates the coupling for the mass eigenstates of $Y Y^\prime$ mixing with two SM particles.
The condition for $X$ is similar as that of $Y$, and  it is also guaranteed by the Z2 symmetry.

In this subsection, we suppose the lightest mass eigenstate of $Y Y^\prime$ mixing in Eq.(\ref{YY'}) as a dark matter candidate, and
calculate the relic density.
So we summarize the relic density constraints that any WIMP candidate has to satisfy.
The interactions of the WIMP with SM particles are deduced from the EBLMSSM, then we study its annihilation rate
and its relic density $\Omega_D$ by the thermal dynamics of the Universe.
The annihilation cross section $\sigma(Y_1 Y_1^* \rightarrow anything)$ should be calculated and can be written as
$\sigma v_{rel}=a+bv_{rel}^2$ in the $Y_1Y_1^*$ center of mass frame. $v_{rel}$ is the twice velocity of $Y_1$ in the $Y_1Y_1^*$ c.m. system frame.
To a good approximation, the freeze-out temperature($T_F$) can be iteratively computed from\cite{dark1}
\begin{eqnarray}
&&x_F=\frac{m_D}{T_F}\simeq\ln[\frac{0.038M_{Pl}m_D(a+6b/x_F)}{\sqrt{g_*x_F}}],
\end{eqnarray}
with $x_F\equiv m_D/T_F$ and $m_D=m_{Y_1}$ representing the WIMP mass. $M_{Pl}=1.22\times10^{19}$ GeV is the Planck mass and $g_*$ is the number
of the relativistic degrees of freedom with mass less than $T_F$. The density of cold non-baryonic matter is
$\Omega_D h^2=0.1186\pm 0.0020$\cite{pdg2016}, whose formula is simplified as
\begin{eqnarray}
\Omega_D h^2\simeq \frac{1.07\times10^9 x_F}{\sqrt{g_*}M_{PL}(a+3b/x_F)\texttt{GeV} }\;.
\end{eqnarray}
To obtain $a$ and $b$ in the $\sigma v_{rel}$, we study the $Y_1Y_1^*$ dominate decay channels whose final states are leptons and light neutrinos:
1. $Y_1Y_1^*\rightarrow Z_L \rightarrow\bar{l}^Il^I$; 2. $Y_1Y_1^*\rightarrow Z_L \rightarrow\bar{\nu}^I\nu^I$;
3. $Y_1Y_1^*\rightarrow \varphi_L \rightarrow\bar{\nu}^I\nu^I$; 4. $Y_1Y_1^*\rightarrow L' \rightarrow\bar{l}^Il^I$;
5. $Y_1Y_1^*\rightarrow N' \rightarrow\bar{\nu}^I\nu^I$.

Using the couplings in Eqs.(\ref{TCLY})(\ref{TCZL})(\ref{TCHL}), we deduce the results of $a$ and $b$
\begin{eqnarray}
&&a=\sum_{l=e,\mu,\tau}\frac{1}{\pi}
|\sum_{i=1}^2\frac{m_{L'_i}}{(m_D^2+m_{L'_i}^2)}\lambda_4W_L^{1i}Z_Y^{11*}\lambda_6U_L^{2i}Z_Y^{21*}|^2
\nonumber\\&&\hspace{0.5cm}+\sum_{\chi^0_{N_\alpha=\nu_e,\nu_\mu,\nu_\tau}}\Big\{\frac{g_L^4(2+L_4)^2 }{8\pi}|(Z_Y^{11*}Z_Y^{11}-Z_Y^{21*}Z_Y^{21})\sum_{I=1}^3\sum_{i=1}^4\frac{1}
{(4m_D^2-m_{\Phi_i}^2)}\nonumber\\&&\hspace{0.5cm}\times(\lambda_{N^c}Z_{N_\nu}^{(I+3)\alpha}
Z_{N_\nu}^{(I+3)\alpha}Z_{\phi_L}^{2i})
(v_LZ^{1i}_{\tilde{\phi}_L}-\bar{v}_LZ^{2i}_{\tilde{\phi}_L}
+\frac{3}{2}v_{NL}Z^{3i}_{\tilde{\phi}_{L}}-\frac{3}{2}\bar{v}_{NL}Z^{4i}_{\tilde{\phi}_{L}})|^2
\nonumber\\&&\hspace{0.5cm}
+\frac{1}{\pi}
|\sum_{i=1}^2\sum_{I=1}^3\frac{m_{N'_i}}{(m_D^2+m_{N'_i}^2)}
\lambda_4Z_{N_{\nu}}^{I\alpha*}W_N^{1i}Z_Y^{11*}\lambda_5Z_{N_{\nu}}^{(I+3)\alpha}U_N^{2i}Z_Y^{21*}|^2\Big\},
\nonumber\\
&&b=\sum_{l=e,\mu,\tau}\frac{7m_D^2 }{24\pi}\frac{g_L^4(2+L_4)^2}{(4m_D^2-m_{Z_L}^2)}
+\sum_{\chi^0_{N_\alpha=\nu_e,\nu_\mu,\nu_\tau}}
\frac{1}{96\pi}
\frac{g_L^4(2+L_4)^2m_D^2}{(4m_D^2-m_{Z_L}^2)^2}\nonumber\\&&\hspace{0.5cm}\times\Big(
7+|\sum_{I=1}^3\Big(Z_{N_\nu}^{I\alpha*}Z_{N_\nu}^{I\alpha}
-Z_{N_\nu}^{(I+3)\alpha*}Z_{N_\nu}^{(I+3)\alpha}\Big)|^2
\Big).
\end{eqnarray}
\section{numerical results}
\subsection{$h^0$ decays and $m_{A^0}, m_{H^0}$}
In this section, we research the numerical results.  For the parameter space, the most strict constraint is that
 the mass of the lightest eigenvector for the mass squared matrix in Eq.(\ref{MCPEHiggs}) is around $125.1$ GeV.
 To satisfy this constraint, we use $m_{h^0}=125.1$ GeV as an input parameter. Therefore, the CP odd Higgs mass should
 meet the following relation.
 \begin{eqnarray}
&&m_{A^0}^2={m_{h^0}^2(m_{Z}^2-m_{h^0}^2+\Delta_{11}+\Delta_{22})-m_{Z}^2
\Delta_{A}+\Delta_{12}^2-\Delta_{11}\Delta_{22}\over -m_{h^0}^2+m_{Z}^2\cos^22\beta
+\Delta_{B}}\;,
\label{Higgs-mass1}
\end{eqnarray}
where
\begin{eqnarray}
&&\Delta_{A}=\sin^2\beta\Delta_{11}+\cos^2\beta\Delta_{22}+\sin2\beta \Delta_{12}
\;,\nonumber\\
&&\Delta_{B}=\cos^2\beta\Delta_{11}+\sin^2\beta\Delta_{22}+\sin2\beta \Delta_{12}\;.
\label{Higgs-mass2}
\end{eqnarray}
To obtain the numerical results, we adopt the following parameters as
\begin{eqnarray}
&&Y_{u_4} = 1.2Y_t,~~~Y_{u_5} = 0.6Y_t,~~~
Y_{d_4}=Y_{d_5} = 2Y_b,~~~ g_B = 1/3,~~~\lambda_u=\lambda_d = 0.5,\nonumber\\&&
A_{u_4} = A_{d_4} = A_{d_5} = A_{e_4} = A_{e_5} =A_{\nu_4} = A_{\nu_5} =1{\rm TeV},~
\lambda_Q = 0.4,~g_L = 1/6,
\nonumber\\&&
m_{\tilde{Q}_4} = m_{\tilde{Q}_5} =m_{\tilde{U}_4} = m_{\tilde{U}_5} = m_{\tilde{D}_4} = m_{\tilde{D}_5} =
m_{\tilde{\nu}_4} = m_{\tilde{\nu}_5} =1{\rm TeV},~Y_{e_5} = 0.6,\nonumber\\&&
\upsilon_{NL} = \upsilon_L =A_b = 3{\rm TeV},~~~~~
\tan\beta_{NL} =\tan\beta_L  = 2,~~~~~ \lambda_L = \lambda_{NL} =\lambda_E = 1,
\nonumber\\&&
m_{\tilde{L}} = m_{\tilde{e}} = 1.4 \delta_{ij}{\rm TeV},~
A_{\tilde{L}} = A_{\tilde{L}'} = 0.5 \delta_{ij}{\rm TeV}~(i,j=1,2,3),~\mu_B = 0.5 {\rm TeV}
, \nonumber\\&&
A_{BQ}=A_{BU} = A_{BD} = \mu_{NL} =A_{LL} = A_{LE} = A_{LN} = 1{\rm TeV},~Y_{\nu_4} = Y_{\nu_5} = 0.1,
\nonumber\\&&
 m_{\tilde{L}_4} = m_{\tilde{L}_5}=
 m_{\tilde{E}_4} = m_{\tilde{E}_5}=m_2=1.5{\rm TeV},~m_{\tilde{D}_3}= 1.2{\rm TeV},~B_4 = L_4 = 1.5. \label{canshu}
\end{eqnarray}
Here $Y_t$ and $Y_b$ are the Yukawa coupling constants of top quark and bottom quark, whose concrete forms are
$Y_t = \sqrt{2} m_t/(\upsilon \sin\beta)$ and $Y_b = \sqrt{2} m_b/(\upsilon \cos\beta)$ respectively.

\begin{figure}[h]
\setlength{\unitlength}{1mm}
\centering
\includegraphics[width=2.9in]{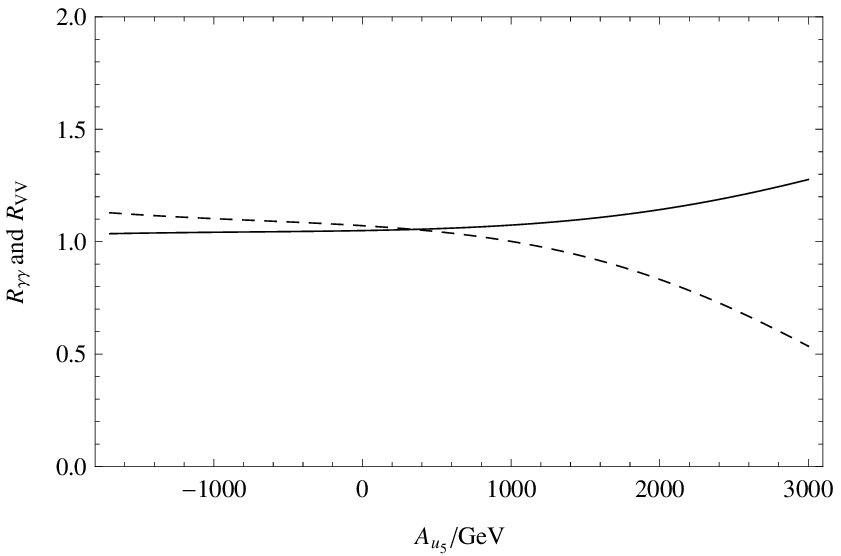}~~~\includegraphics[width=2.9in]{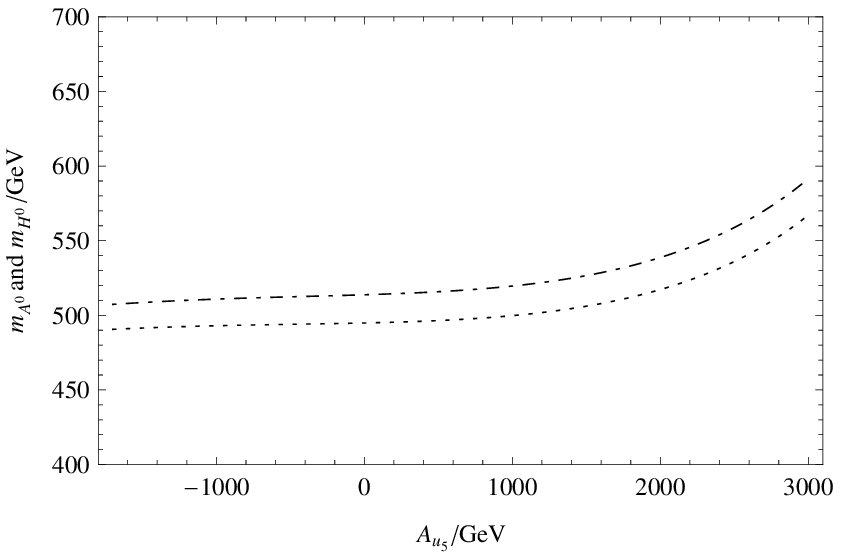}
\caption[]{The results versus $A_{u_5}$ are shown.
$R_{\gamma\gamma}$ (solid line) and $R_{VV}$ (dashed line) are in the left diagram. $m_{A^0}$ (dot-dashed line)
and $m_{H^0}$ (dotted line) are in the right diagram.}\label{Au5tu}
\end{figure}

To embody the exotic squark corrections, we calculate the results versus $A_{u_5}$ which has relation with the mass squared matrix of exotic squark.
In the left diagram of FIG.\ref{Au5tu}, $R_{\gamma\gamma}$ and $R_{VV}$ versus $A_{u_5}$ are plotted by the solid line and dashed line
 respectively with $m_{\tilde{Q}_3}=m_{\tilde{U}_3}=1.2{\rm TeV},~\tan\beta =1.4, ~A_t=1.7{\rm TeV},\upsilon_B =3.6{\rm TeV},~\mu =-2.4{\rm TeV},~\tan\beta_B =1.5$ and $Y_{e_4}=0.5$. In the left diagram of FIG.\ref{Au5tu}, the solid line($R_{\gamma\gamma}$)
 and dashed line ($R_{VV}$) change weakly with the $A_{u_5}$.
 When $A_{u_5}$ enlarges, $R_{\gamma\gamma}$ is the increasing function and $R_{VV}$ is the decreasing function.
 During the $A_{u_5}$ region $(-1700\sim1000)$ GeV, both $R_{\gamma\gamma}$ and $R_{VV}$ satisfy the experiment limits.  The dot-dashed line(dotted line) in the right diagram  denotes
 the Higss mass $m_A^0(m_H^0)$ varying with $A_{u_5}$. The dot-dashed line and dotted line
 increase mildly with $A_{u_5}$. The value of $m_A^0$ is a little bigger than 500 GeV, while the value of $m_H^0$ is very near 500 GeV.

\begin{figure}[h]
\setlength{\unitlength}{1mm}
\centering
\includegraphics[width=2.9in]{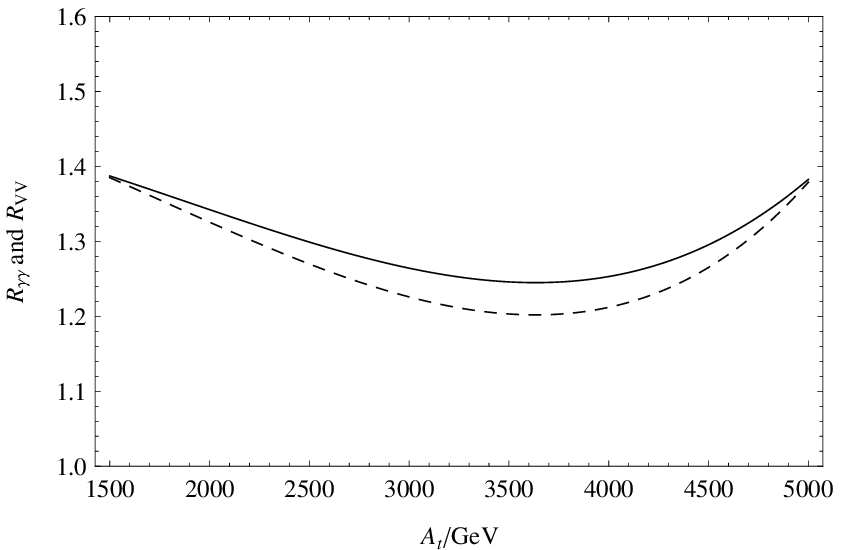}~~~\includegraphics[width=2.9in]{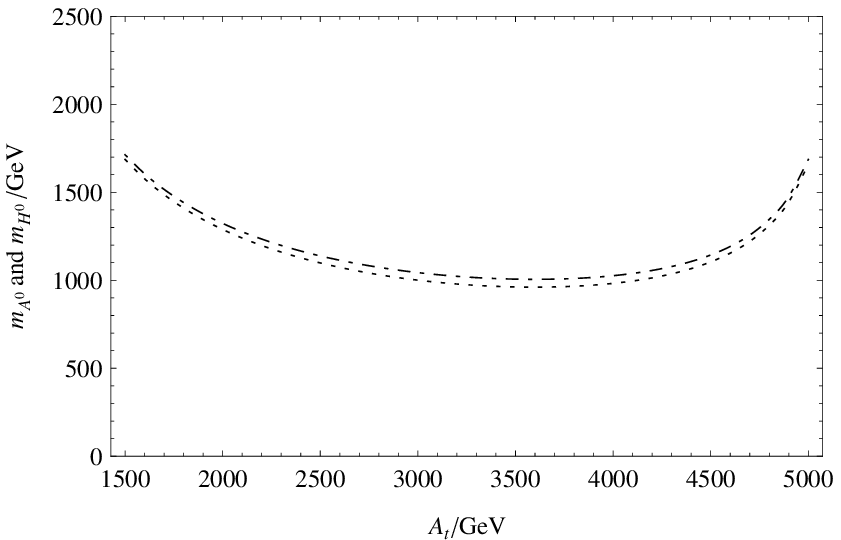}
\caption[]{The results versus $A_{t}$ are shown.
$R_{\gamma\gamma}$ (solid line) and $R_{VV}$ (dashed line) are in the left diagram. $m_{A^0}$ (dot-dashed line)
and $m_{H^0}$ (dotted line) are in the right diagram.}\label{NAttu}
\end{figure}

  For the squark, we assume the first and second generations are heavy, so they are neglected.
 The scalar top quarks are not heavy, and their contributions are considerable.
 $A_t$ is in the mass squared matrix of scalar top quark influencing the mass and mixing.
 The effects from $A_t$ to the ratios $R_{\gamma\gamma}$, $R_{VV}$, Higgs masses $m_{A^0}$  and $m_{H^0}$ are of interest.
 As $m_{\tilde{Q}_3}=2.4{\rm TeV},~ m_{\tilde{U}_3}=1.2{\rm TeV},
 ~\tan\beta =\tan\beta_B =2.15,~\upsilon_B =4.1{\rm TeV},~\mu =-2.05{\rm TeV},
 ~Y_{e_4}=0.5$ and $ A_{u_5}=1 {\rm TeV}$. $R_{\gamma\gamma}$ (solid line) and $R_{VV}$ (dashed line) versus $A_t$ are shown in the left diagram of FIG.\ref{NAttu}.
 While the right diagram of FIG.\ref{NAttu} gives out the Higgs masses  $m_{A^0}$ (dot-dashed line) and $m_{H^0}$ (dotted line).
  In the $A_t$ region $(2\sim4.8)$ TeV, the $R_{\gamma\gamma}$ varies from 1.25 to 1.34. At the same time,
 the $R_{VV}$ is in the range $(1.2\sim1.38)$.
 The dot-dashed line and dotted line are very near. In the $A_t$ region $(3000\sim4000)$ GeV, the masses of Higgs $A^0$ and $H^0$ are around 1000 GeV.
 In this parameter space, the allowed biggest values of $A^0$ and $H^0$ masses can almost reach 1350 GeV.

\begin{figure}[h]
\setlength{\unitlength}{1mm}
\centering
\includegraphics[width=2.9in]{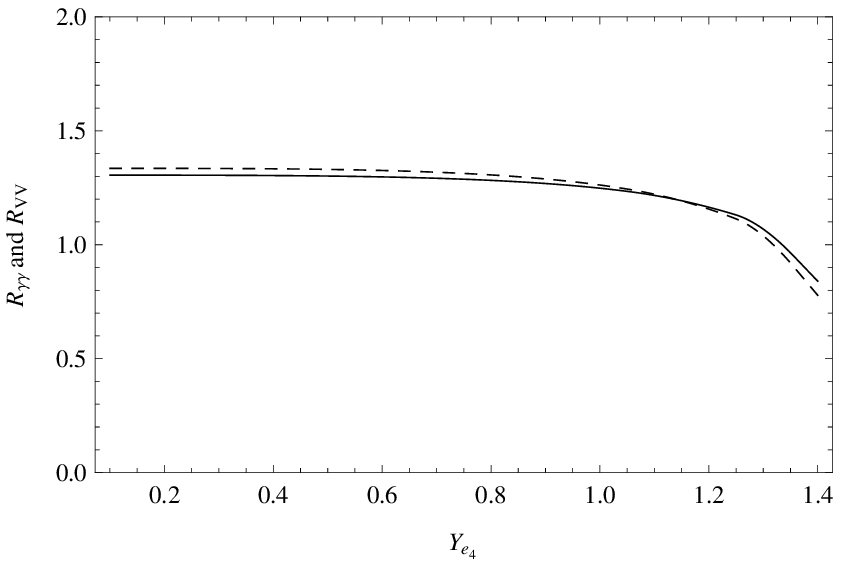}~~~\includegraphics[width=2.9in]{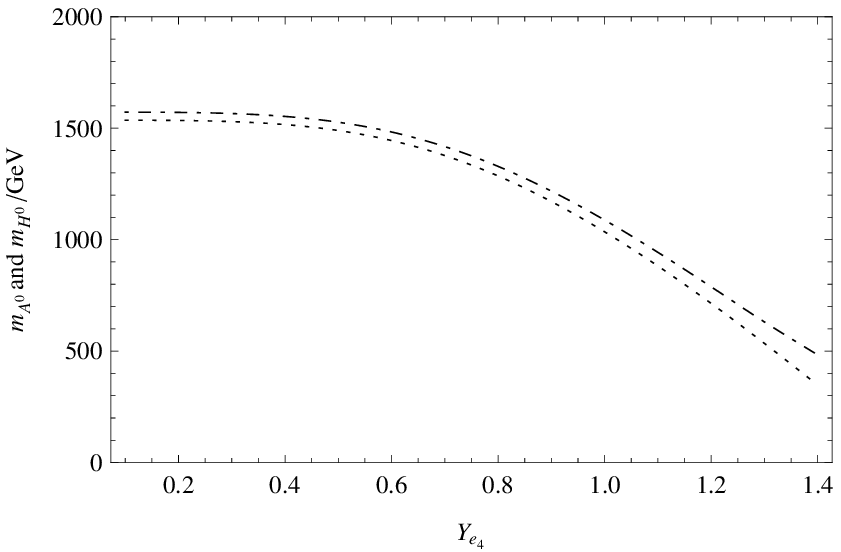}
\caption[]{The results versus $Y_{e_4}$ are shown.
$R_{\gamma\gamma}$ (solid line) and $R_{VV}$ (dashed line) are in the left diagram. $m_{A^0}$ (dot-dashed line)
and $m_{H^0}$ (dotted line) are in the right diagram.}\label{NYe4tu}
\end{figure}

$Y_{e_4}$ is the Yukawa coupling constant that can influence the mass matrix of exotic lepton and exotic slepton.
We use $m_{\tilde{Q}_3}= m_{\tilde{U}_3}=1.2{\rm TeV},~\tan\beta =2.3,~\tan\beta_B =1.77,
~A_t=1.7{\rm TeV},~\upsilon_B =5.43{\rm TeV},~\mu =-2.64{\rm TeV}
,~ A_{u_5}=1 {\rm TeV}$ and obtain the results versus $Y_{e_4}$ in the FIG.\ref{NYe4tu}.
In the left diagram, the  $R_{\gamma\gamma}$ (solid line) and $R_{VV}$ (dashed line) are around 1.3 and their changes
are small during the $Y_{e_4}$ range $(0.05\sim1)$.
One can see that in the right diagram $m_{A^0}$ (dot-dashed line) and $m_{H^0}$ (dotted line) possess same behavior versus $Y_{e_4}$.
They are both decreasing functions of $Y_{e_4}$ and vary from 1500GeV to 500 GeV.
In general, $Y_{e_4}$ effect to the Higgs masses $m_{A^0}$ and $m_{H^0}$ is obvious.

\begin{figure}[h]
\setlength{\unitlength}{1mm}
\centering
\includegraphics[width=2.9in]{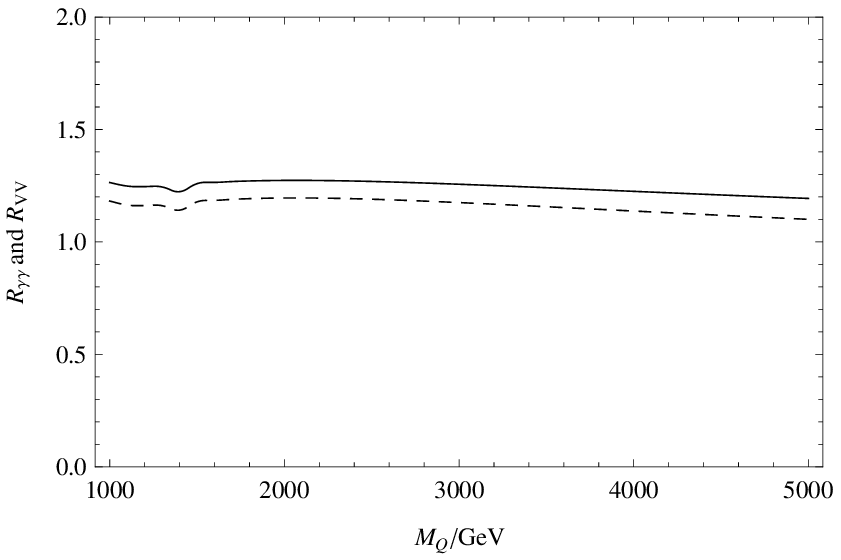}~~~\includegraphics[width=2.9in]{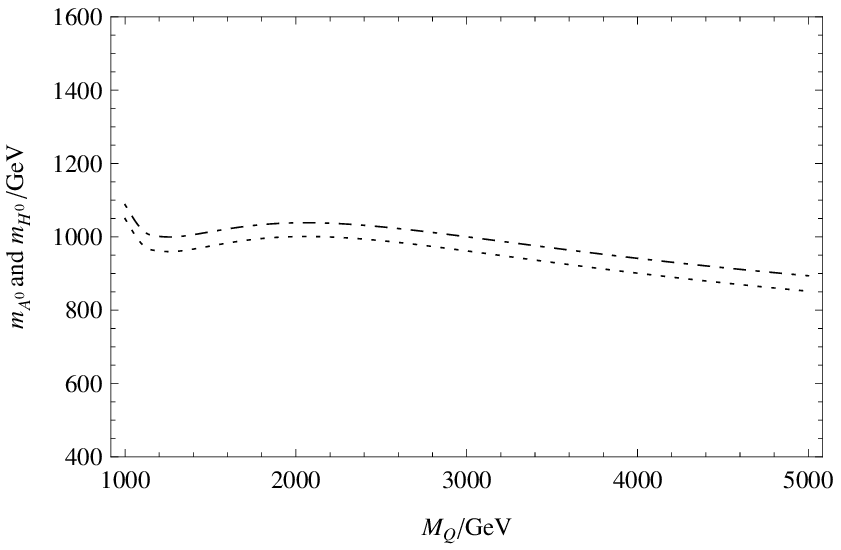}
\caption[]{The results versus $M_{Q}$ are shown.
$R_{\gamma\gamma}$ (solid line) and $R_{VV}$ (dashed line) are in the left diagram. $m_{A^0}$ (dot-dashed line)
and $m_{H^0}$ (dotted line) are in the right diagram.}\label{NMQtu}
\end{figure}

$m_{\tilde{Q}_3}$ and $m_{\tilde{U}_3}$ are the diagonal elements of the squark mass squared matrix,
and they should affect the results. Supposing $m_{\tilde{Q}_3}= m_{\tilde{U}_3}=M_Q,~\tan\beta =2.1,
~\tan\beta_B =2.24,~A_t=1.7{\rm TeV},~\upsilon_B =3.95{\rm TeV},~\mu =-1.9{\rm TeV}
,~Y_{e_4}=0.6,~ A_{u_5}=1 {\rm TeV}$, we calculate the results versus $M_Q$ and plot the diagrams in the FIG.\ref{NMQtu}.
It shows that in this figure the solid line, dashed line, dotted line and dot-dashed line are all stable.
$R_{\gamma\gamma}$ and $R_{VV}$ are around 1.2. At the same time $m_{A^0}$ and $m_{H^0}$ are about 1 TeV.

\subsection{scalar dark matter $Y_1$}
\begin{figure}[h]
\setlength{\unitlength}{1mm}
\centering
\includegraphics[width=2.9in]{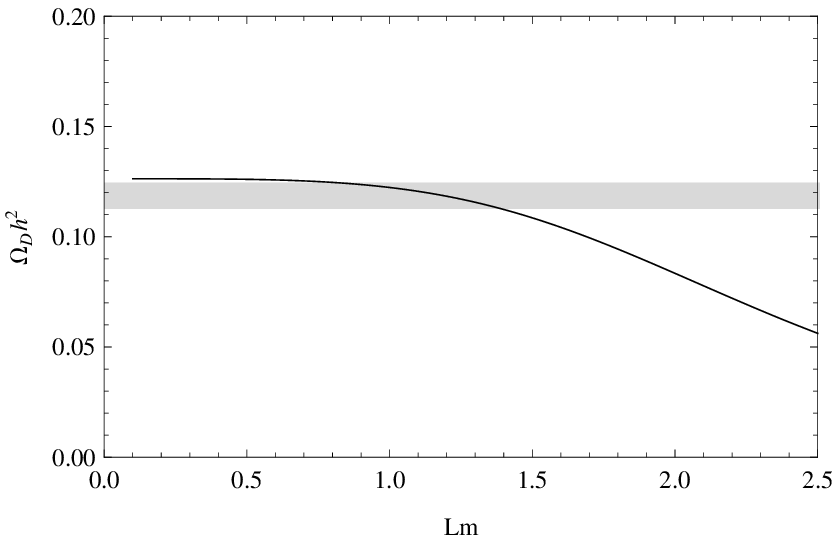}~~~\includegraphics[width=2.9in]{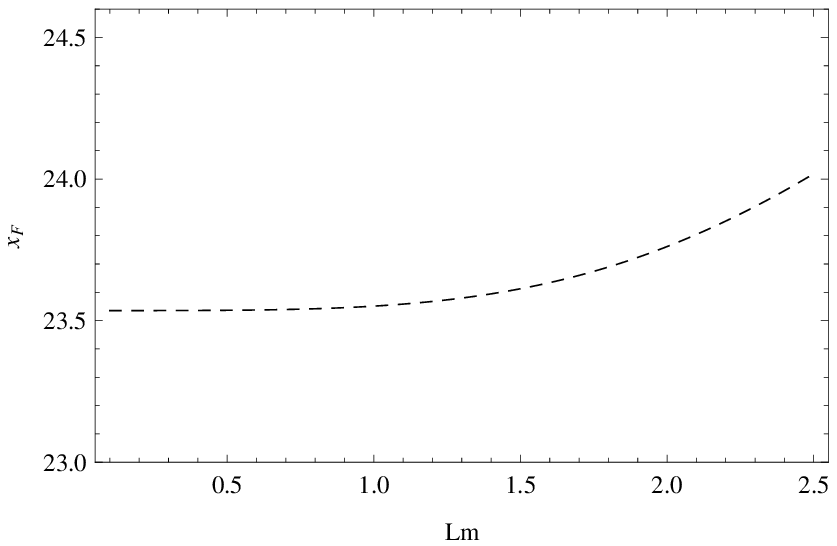}
\caption[]{The relic density and $x_F$ versus $Lm$.}\label{darktu1}
\end{figure}

Here, we suppose $Y_1$ as a scalar dark matter candidate. 
In Ref.\cite{pdg2016} the density of cold non-baryonic matter is
$\Omega_D h^2=0.1186\pm 0.0020$.
To obtain the numerical results of dark matter relic density, for consistency the used parameters in this subsection are of the same values as in Eq.(\ref{canshu})
if they are supposed. Therefore, we just show the values of the parameters beyond Eq.(\ref{canshu}). These parameters are taken as
\begin{eqnarray}
&&~\mu_Y=1500{\rm GeV},~~~~~ \lambda_5=1,~~~~~
\mu_L =B_L = B_{NL} = 1{\rm TeV},~~~~~\tan\beta=1.4,\nonumber\\&&
B_Y=940{\rm GeV},~~~~~m_{\Phi_L}^2=m_{\varphi_L}^2= m_{\Phi_{NL}}^2= m_{\varphi_{NL}}^2= 3 {\rm TeV}^2,~~~~Y_{e_4}=0.5.\label{darkparameter}
\end{eqnarray}
With the relation $\lambda_4=\lambda_6=Lm$, we study relic density $\Omega_D$ and $x_F$ versus $Lm$ in the FIG.\ref{darktu1}.
In the right diagram of FIG.\ref{darktu1}, the grey area is the experimental results in 3 $\sigma$ and
the solid line representing $\Omega_Dh^2$ turns small with the increasing $Lm$.
During the $Lm$ region $(0.7\sim1.4)$, $\Omega_Dh^2$ satisfies the experiment bounds of dark matter relic density.
$x_F$ is stable and in the region $(23.5 \sim 24)$.

\begin{figure}[h]
\setlength{\unitlength}{1mm}
\centering
\includegraphics[width=2.9in]{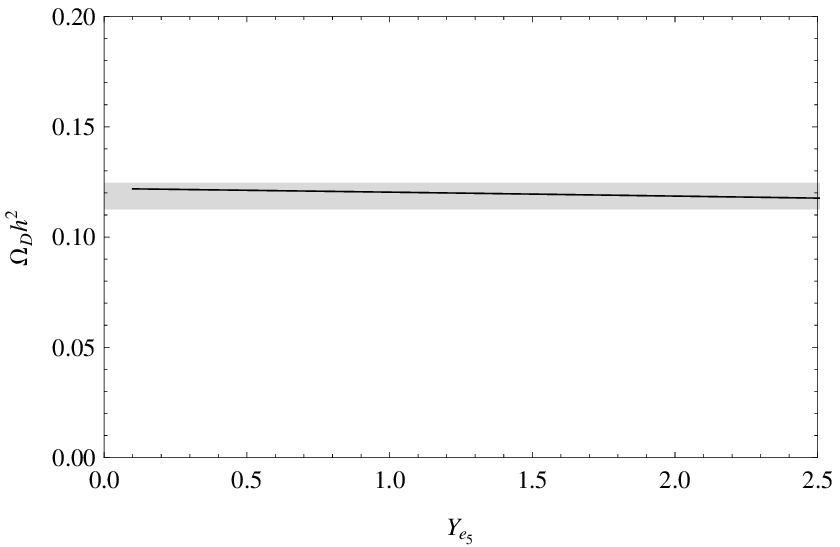}~~~\includegraphics[width=2.9in]{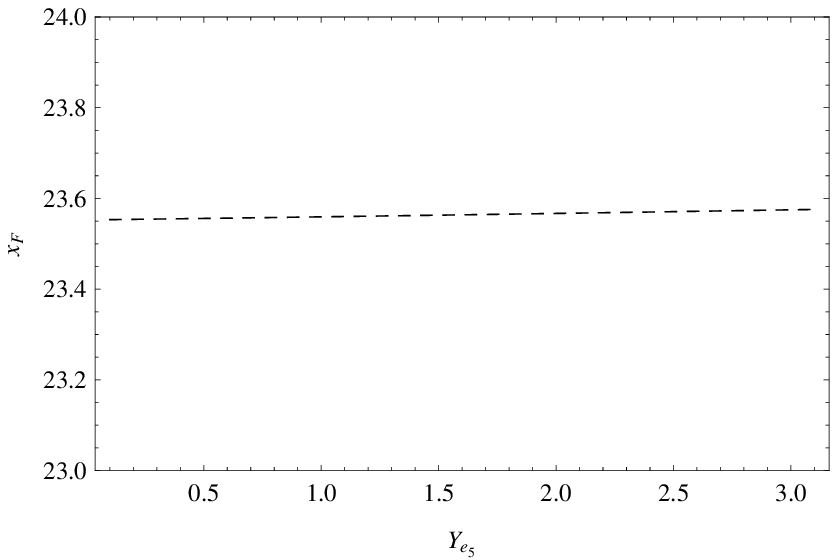}
\caption[]{The relic density and $x_F$ versus $Y_{e_5}$.}\label{darktu2}
\end{figure}

 Taking $Y_{e_4}=1.3, \lambda_4=\lambda_6=1$ and the other parameters being same as Eq.(\ref{darkparameter}) condition,
 we plot the relic density($x_F$) versus $Y_{e_5}$ in the left (right) diagram of the FIG.\ref{darktu2}.
 In this parameter space, during $Y_{e_5}$ region $(0.1\sim2.5)$, our theoretical results satisfy the
 relic density bounds of dark matter, and
 $x_F$ is very near 23.55. Generally speaking, both the solid line and dashed line are very stable.

\section{discussion and conclusion}

   Considering the light exotic lepton in BLMSSM, we add exotic Higgs superfields
    $\Phi_{NL}$ and $\varphi_{NL}$ to BLMSSM in order to make the exotic leptons heavy.
Light exotic leptons may be excluded by the experiment in the future. On the other hand, heavy exotic leptons
should not be stable. So we also introduce
the superfields $Y$ and $Y'$ to make exotic leptons decay quickly. The lightest mass eigenstate of $Y$ and $Y'$ mixing mass matrix can be a dark matter candidate.
Therefore, the exotic leptons are heavy enough to decay to SM leptons and Y at tree level. We call this extended BLMSSM as EBLMSSM,
where the mass matrices for the particles are deduced and compared with those in BLMSSM. Different from BLMSSM, the exotic sleptons of 4 and 5
generations mix together forming $4\times4$ mass squared matrix. EBLMSSM has more abundant content than BLMSSM for the lepton physics.

To confine the parameter space of EBLMSSM, we study the decays $h^0\rightarrow \gamma\gamma$ and $h^0\rightarrow VV, V=(Z,W)$. 
The CP even Higgs masses $m_{h^0}, m_{H^0}$ and
CP odd Higgs mass $m_A^0$ are researched. In the numerical calculation, to keep $m_{h^0}=125.1$ GeV, we use it as an input parameter.
In our used parameter space, the values of $R_{\gamma\gamma}$ and $R_{VV}$ both meet the experiment limits. The CP odd Higgs mass $m_{A^0}$
is a little heavier than the CP even Higgs mass $m_{H^0}$.  Generally speaking, both $m_{A^0}$ and $m_{H^0}$ are in the region $(500\sim1500)$ GeV.
Based on the supposition that the lightest mass eigenstate $Y_1$ of $Y$ and $Y'$ mixing possesses the character of cold dark matter, we research
the relic density of $Y_1$. In our used parameter space, $\Omega_Dh^2$ of $Y_1$ can match the experiment bounds.
EBLMSSM has a bit more particles and parameters than those in BLMSSM. Therefore, EBLMSSM possesses stronger adaptive capacity
 to explain the experiment results and some problems in the theory.  In our later work, we shall study the EBLMSSM and confine its parameter space
 to move forward a single step.

{\bf Acknowledgments}

Supported by the Major Project of NNSFC (No. 11535002, No. 11605037, No. 11705045),
the Natural Science Foundation of Hebei province with Grant
No. A2016201010 and No. A2016201069, and the Natural Science Fund of
Hebei University with Grants No. 2011JQ05 and No. 2012-
242, Hebei Key Lab of Optic-Electronic Information and
Materials, the midwest universities comprehensive strength
promotion project. At last, thanks Dr. Tong Li and Dr. Wei Chao very much for useful discussions of dark matter.

\end{document}